\documentclass{aa}

\usepackage{graphicx}
\usepackage{txfonts}
\usepackage{hyperref}
\usepackage{natbib}
\bibpunct{(}{)}{;}{a}{}{,}

\usepackage{xcolor}
\definecolor{dark-red}{rgb}{0.9,0.0,0.0}
\definecolor{dark-blue}{rgb}{0.15,0.15,0.9}
\definecolor{dark-green}{rgb}{0.15,0.8,0.15}
\definecolor{medium-blue}{rgb}{0,0,0.9}
\hypersetup{
    colorlinks, linkcolor=red,
    citecolor={dark-blue} , urlcolor={medium-blue}
}
\newcommand{\teff}{$T_{\text{eff}}$\,}
\newcommand{\logg}{$\log$g\,}
\newcommand{\FeI}{Fe\,{\sc i}\,}
\newcommand{\FeII}{Fe\,{\sc ii}\,}
\newcommand{\feh}{$[$Fe/H$]$\,}
\newcommand{\vt}{$\xi_t$\,}
\newcommand{\angstrom}{\text{\normalfont\AA}}
\newcommand{\kms}{km s$^{-1}$\,}
\newcommand{\msun}{M$_\odot$\,}

\newcommand{\loggtri}{$\log$g$_{\text{tri}}$\,}

\begin{document}

   \title{SPECIES II. Stellar parameters of the EXPRESS program\\ giant star sample}


   \author{M.~G.~Soto
          \inst{1}
          \and
          M.~I.~Jones\inst{2,3}
          \and
          J.~S.~Jenkins\inst{4}
          }

   \institute{School of Physics and Astronomy, Queen Mary University London, 327 Mile End Road, London E1 4NS, UK\\\email{m.soto@qmul.ac.uk}
         \and
         European Southern Observatory, Alonso de C\'ordova 3107, Vitacura, Casilla 19001, Santiago, Chile
         \and
         Instituto de Astronom\'ia, Universidad Cat\'olica del Norte, Angamos 0610, 1270709, Antofagasta, Chile
         \and
         Departamento de Astronom\'ia, Universidad de Chile, Camino El Observatorio 1515, Las Condes, Santiago, Chile
             }

   \date{Received XX; accepted XX}

 
  \abstract
   {As part of the search for planets around evolved stars, we can understand planet populations around significantly higher-mass stars than the Sun on the main sequence. This population is difficult to study any other way, particularly with radial-velocities to measure planet masses and orbital mechanics, since the stars are too hot and rotate too fast to present the quantity of narrow stellar spectral lines necessary to measure velocities at the few m/s level.}
   {Our goal is to estimate stellar parameters for all of the giant stars from the EXPRESS project, which aims to detect planets orbiting evolved stars, and study their occurrence rate as a function of stellar mass.}
   {We analyse high resolution echelle spectra of these stars, and compute their atmospheric parameters by measuring the equivalent widths for a set of iron lines, using an updated method implemented during this work. Physical parameters, like mass and radius, are computed by interpolating through a grid of stellar evolutionary models, following a procedure that carefully takes into account the post-main sequence evolutionary phases.
   The atmospheric parameters, as well as photometric and parallax data, are used as constraints during the interpolation process. Probabilities of the star being in the red giant branch (RBG) or the horizontal branch (HB) are estimated from the derived distributions.}
   {We obtain atmospheric and physical stellar parameters for the whole EXPRESS sample, which comprises a total of 166 evolved stars. 
   We find that
   101 of them are most likely first ascending the RGB phase, while 65 of them have already reached the HB phase. The mean derived mass is $1.41 \pm 0.46\,M_{\odot}$ and $1.87 \pm 0.53\,M_{\odot}$ for RGB and HB stars, respectively. \newline \indent
   To validate our method, we compared our derived physical parameters with interferometry and asteroseismology studies. In particular, when comparing to stellar radii derived from interferometric angular diameters we find: $\Delta R_{\text{inter}} = -0.11\, R_{\odot}$, which corresponds to a 1.7\% difference. Similarly, when comparing with asteroseismology we obtain the following results: $\Delta \log{\text{g}} = 0.07$ cgs (2.4\%), $\Delta R = -0.12\, R_{\odot}$ (1.5\%), $\Delta M = 0.08\, M_{\odot}$ (6.2\%) and $\Delta \text{age} = -0.55$ Gyr (11.9\%). 
   Additionally, we compared our derived atmospheric parameters with previous spectroscopic studies. We find the following results: $\Delta$\teff = 22 K (0.5\%), $\Delta$\logg = -0.03 (1.0 \%) and $\Delta$\feh = -0.04 dex (2\%).
   We also find a mean systematic difference in the mass with respect to those presented in the EXPRESS original catalogue of $\Delta M = -0.28 \pm 0.27\, M_{\odot}$, corresponding to a systematic mean difference of 16\%. For the rest of the atmospheric and physical parameters we find good agreement between the original catalogue and the results presented here.
   Finally, we find excellent agreement between the spectroscopic and trigonometric \logg values, showing the internal consistency and robustness of our method.}
   {We show that our method, which includes a re-selection of iron lines and changes in the interpolation of evolutionary models, as well as Gaia parallaxes and newer extinction maps, can greatly improve the estimates of stellar parameters for giant stars compared to those presented in our previous work. This method also results in smaller mass estimates, an issue that has been found in results for giant stars from spectroscopy in the literature. The results provided here will improve the physical parameter estimates of planetary companions found orbiting these stars, and give us insights into their formation and the effect of stellar evolution on their survival.}

   \keywords{stars: fundamental parameters -- stars: horizontal-branch -- techniques: spectroscopic  }

   \maketitle
%

\section{Introduction}

During the past 25 years, more than 4200\footnote{\url{http://exoplanet.eu}} planets orbiting stars outside of our Solar System have been found, changing our views regarding planet formation and evolution. 
Thanks to numerous efforts to characterise the stars where these systems were found (and where no planet has been discovered yet), some correlations have come to light. 
One of them is the so-called planet-metallicity relation, in which there appears to be a positive correlation between giant planet fraction and stellar metallicity \citep[e.g. ][]{Fischer2005,Jenkins2017}, and points to the core accretion model for planet formation as the most probable responsible mechanism \citep[e.g. ][]{Ida2004}. The planet-metallicity relation has been well established for main-sequence stars with stellar masses $M < 1.5\, M_{\odot}$, but it is still uncertain for more massive stars, with mixed results coming from different studies \citep{Jones2016, Reffert2015, Mortier2013a, Maldonado2013, Ghezzi2010, Hekker2007, Pasquini2007}.
One reason for this is that intermediate-mass stars ($M > 1.5\, M_{\odot}$), due to their high temperatures during the main sequence (\teff $\gtrsim$ 6000 K, mainly A-F spectral type), and their high rotational velocities ($\sim 140$ \kms for a $M \sim 1.5\, M_{\odot}$ star; \citealt{Royer2007}), have fewer and broader absorption lines than cooler objects, making them unfavourable candidates in radial velocity (RV) surveys as the measurement of radial velocities is more challenging \citep{Galland2005, Lagrange2009, Borgniet2019}.
One way to access intermediate-mass stars for planet detection is to study them as they evolve off the main sequence.
As a star evolves off the main sequence, its temperature decreases (more absorption lines in its spectra), as well as its rotational velocity \citep{Schrijver1993}, resulting in narrower line profiles. These two effects translate into more precise RV measurements, allowing m/s RV precision to be achieved.
As a result, during the last 20 years, multiple RV surveys have focused on evolved stars \citep[]{Frink2001, Setiawan2003, Sato2005, Hatzes2005, Niedzielski2007, Johnson2007, Jones2011, Wittenmyer2011}. 
\newline \indent
The precise determination of the stellar parameters have direct repercussions in planetary studies, and therefore large programs generally set out to calculate the stellar parameters early in the program \citep[e.g. ][]{Santos2003, Valenti2005, Jenkins2008}. The physical characteristics (mass and radius) of a planet depend on the characteristics of the host star, and uncertainties in these values can be translated into uncertainties in planetary density and composition. 
The physical parameters for a main sequence star (like its mass) can be estimated from their position in the HR diagram (or colour-magnitude diagram), as evolutionary tracks for different masses are well separated from each other (for a given metallicity). That is not the case for evolved stars.
Evolutionary tracks after the main sequence are degenerate in these diagrams, which means that stars with different masses and evolutionary states occupy very similar positions, making the solutions completely different depending on the chosen evolutionary state. This can be illustrated in Fig.~\ref{fig:HRdiagram}, where it can be seen that the Red Giant Branch (RGB) for a $2\,M_{\odot}$ star lies very close to the Horizontal Branch (HB, sometimes also referred to as the Red Clump) of a $1\,M_{\odot}$ star. For a given position in the HR diagram, the RGB solution will be more massive than the HB one. This can lead, for example, to the overestimation of the mass of a star's sample if the stars have been incorrectly assigned to the RGB \citep{Takeda2015, Stock2018}.
If we correctly determine the star's evolutionary state, we can then estimate its position in the main sequence and then reconstruct the changes it went through up to the current state \citep{Villaver2014}. 
\newline \indent
The EXoPlanets aRound Evolved StarS \citep[EXPRESS, ][hereafter J11]{Jones2011} program studies a sample of 166 evolved stars in the southern hemisphere, looking for planetary companions using the radial velocity method. 
The program has already detected 19 planetary companions orbiting 17 giants stars \citep{Jones2013, Jones2014, Jones2015a, Jones2015b, Jones2016, Jones2017, Jones2020}. Some targets in the sample also have companions detected by using combined datasets and also by other groups targeting common targets \citep{Fischer2009, Johnson2011, Sato2012, Trifonov2014, Wittenmyer2016c, Wittenmyer2016b, Wittenmyer2017}.
The aim of EXPRESS is to establish the rate of close-in planets ($P \leq 150$ days) around RGB and HB stars.
There is observational evidence of a lack of close-in planets around evolved stars \citep{Johnson2007, Sato2008b, Dollinger2009, Jones2014, Jones2020}. Different possible scenarios have been proposed to explain this: tidal interaction between the star and planet after the star leaves the main-sequence, resulting in planetary engulfment \citep{Villaver2009, Kunitomo2011, Villaver2014}, an inherent low gas giant formation efficiency at short orbital separations around intermediate-mass stars \citep{Currie2009}, and a rapid disk dissipation that prevents gas giants from migrating to close-in orbits \citep{Ribas2015}.

There has been ongoing debate over whether the mass of "massive" evolved planetary-host stars has been overestimated in the literature, and whether these stars are actually the evolved counterparts of F-G stars, as opposed to massive A-stars \citep{Lloyd2011, Lloyd2013, Schlaufman2013}. In such a scenario, planet engulfment would likely result in a lack of close-in planets orbiting evolved stars, as we know of many close-in planets orbiting Sun-like stars on the main sequence \citep{Schlaufman2013}.
This mass overestimation has also been found in studies of other datasets. For example, \citet{Stock2018} recomputed the masses of the targets from the Lick Planet Search \citep{Frink2001} and compared them to those presented in \citet{Reffert2015}. They found a $\Delta$\,$M_\star$ distribution with a mean value of -0.12\,M$_{\odot}$ and standard deviation of  0.47\,M$_{\odot}$.
The debate regarding the true mass distribution of giant stars in planet search surveys has continued \citep{Johnson2013, Sousa2015, Ghezzi2018}, and highlights the importance of comparing the mass estimates with results from other mass determination methods, for example, asteroseismology \citep{Serenelli2020}.

In this paper we discuss an extension of the SPECIES code \citep[hereafter SJ18]{Soto2018}, first used on main sequence stars, towards the giant phase. We present an update on the stellar parameters and evolutionary status for the EXPRESS sample, from what was presented in J11. We also compare our results with asteroseismic and interferometric studies, and find that they agree within the uncertainties.
The paper is organised as follows: in Sect.~\ref{sec:sample} we present the EXPRESS stars sample; in Sect.~\ref{sec:method} we describe the method used for the stellar parameters computation; in Sect.~\ref{sec:results} we present and discuss our results, and how they compare to other works in the literature; finally, in Sect.~\ref{sec:summary} we present our conclusions and summary.


\section{The sample}\label{sec:sample}

\begin{figure}
\centering
\includegraphics[width=\hsize]{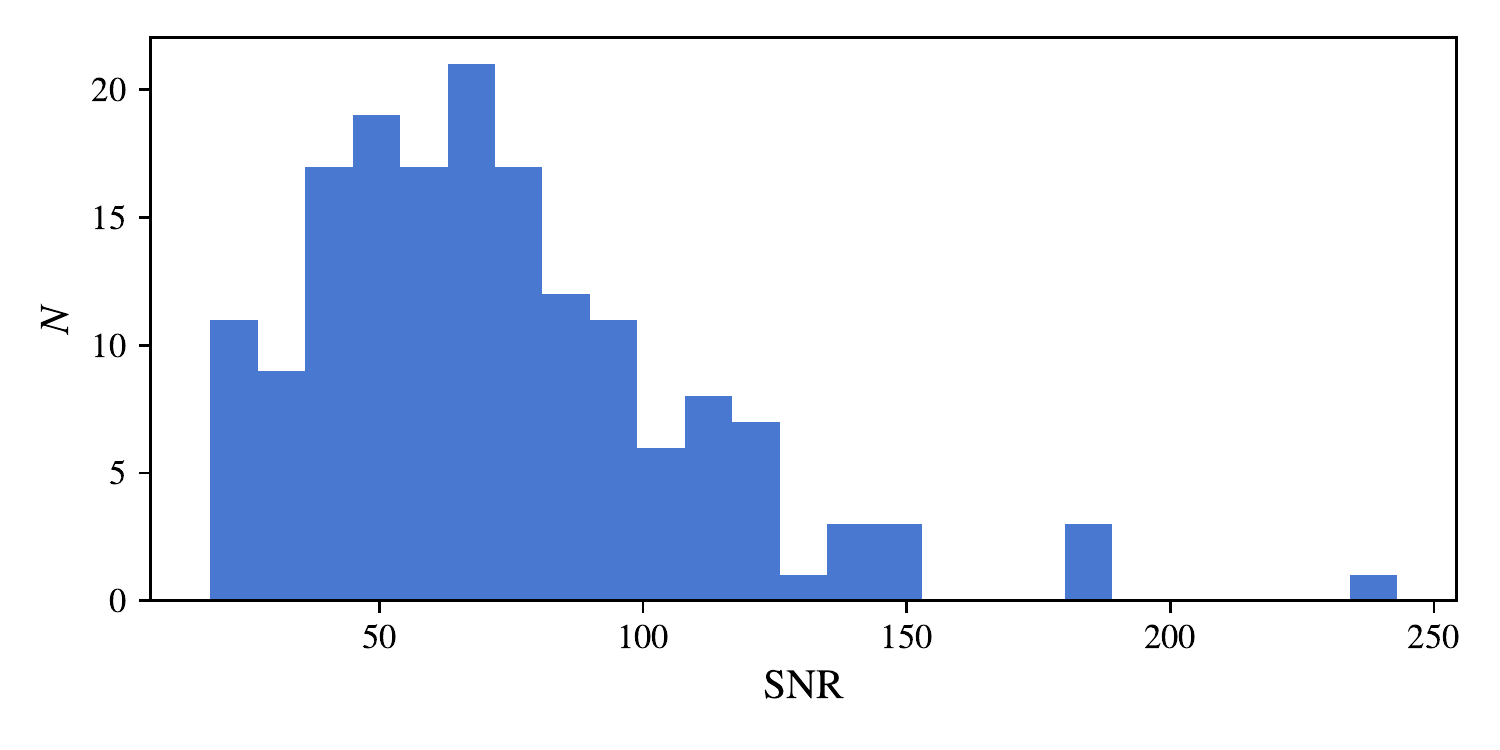}
\caption{SNR distribution of our spectra for the EXPRESS sample.}
\label{fig:SNR}
\end{figure}

\begin{table}
\caption{Spectral resolution $R = \lambda/\Delta\lambda$ for the spectrographs used in this work.}
    \label{tab:resolution}
    \centering
    \begin{tabular}{c | c  c  c  c}
    \hline\hline
         & FEROS & HARPS & CHIRON & FIDEOS  \\
         \hline
        $R$ & 48000 & 120000 & 79000 & 43000\\
        \hline
    \end{tabular}
\end{table}

The EXPRESS program sample consists of 166\footnote{In the sample presented in J11 there were two missing stars. } evolved stars, observable from the southern hemisphere ($\delta$ $\le$ +20 deg). The stars were selected according to the following selection criteria: 0.8 $\le$ B-V $\le$ 1.2, -4.0 $\le$ M$_V$ $\le$ 0.5 and V $\le$ 8.0. These criteria allowed us to include relatively bright first-ascending red giant branch (RGB) stars, as well as clump giants, with intrinsic RV noise below $\sim$ 20 m\,s$^{-1}$ \citep{Sato2005,Hekker2006}. \newline
In addition, we imposed a photometric variability $\le$ 0.02 mag and parallax precision better than $\sim$ 14\%, from the Hipparcos\, catalogue \citep{Perryman1997}. Finally, we excluded stars with known (sub)stellar companions from the sample. \newline \indent
As part of the EXPRESS project, we have collected multi-epoch high-resolution spectroscopic data for all 166 stars, using the Fiber-fed Extended Range Optical Spectrograph \citep[FEROS;][]{Kaufer1999}, mounted on the MPG/ESO 2.2m telescope at La Silla Observatory, the decommissioned Fiber Echelle spectrograph (FECH) mounted on the 1.5m telescope at CTIO, and CHIRON \citep{Tokovinin2013}, which replaced FECH at the 1.5m telescope. Additional spectroscopic data were obtained using the FIber Dual Echelle Optical Spectrograph \citep[FIDEOS;][]{Vanzi2018}, at the 1\,m UCN telescope at La Silla, to characterise binary companions \citep{Bluhm2016} and we also retrieved HARPS archival data available for some of our targets. The spectral resolution of all the instruments is listed in Table~\ref{tab:resolution}.  \newline \indent
The reduction of the FEROS and FIDEOS data was performed using the CERES pipeline \citep{Brahm2017}. Similarly, the CHIRON data were reduced with the Yale pipeline \citep{Tokovinin2013}. In the case of HARPS, we used the ESO data reduction system.
\newline \indent
The SNR distribution of our spectra is shown in Fig.~\ref{fig:SNR}, with a median of 69 and standard deviation of 36.

\section{Method}\label{sec:method}

\begin{figure}
\centering
\includegraphics[width=\hsize]{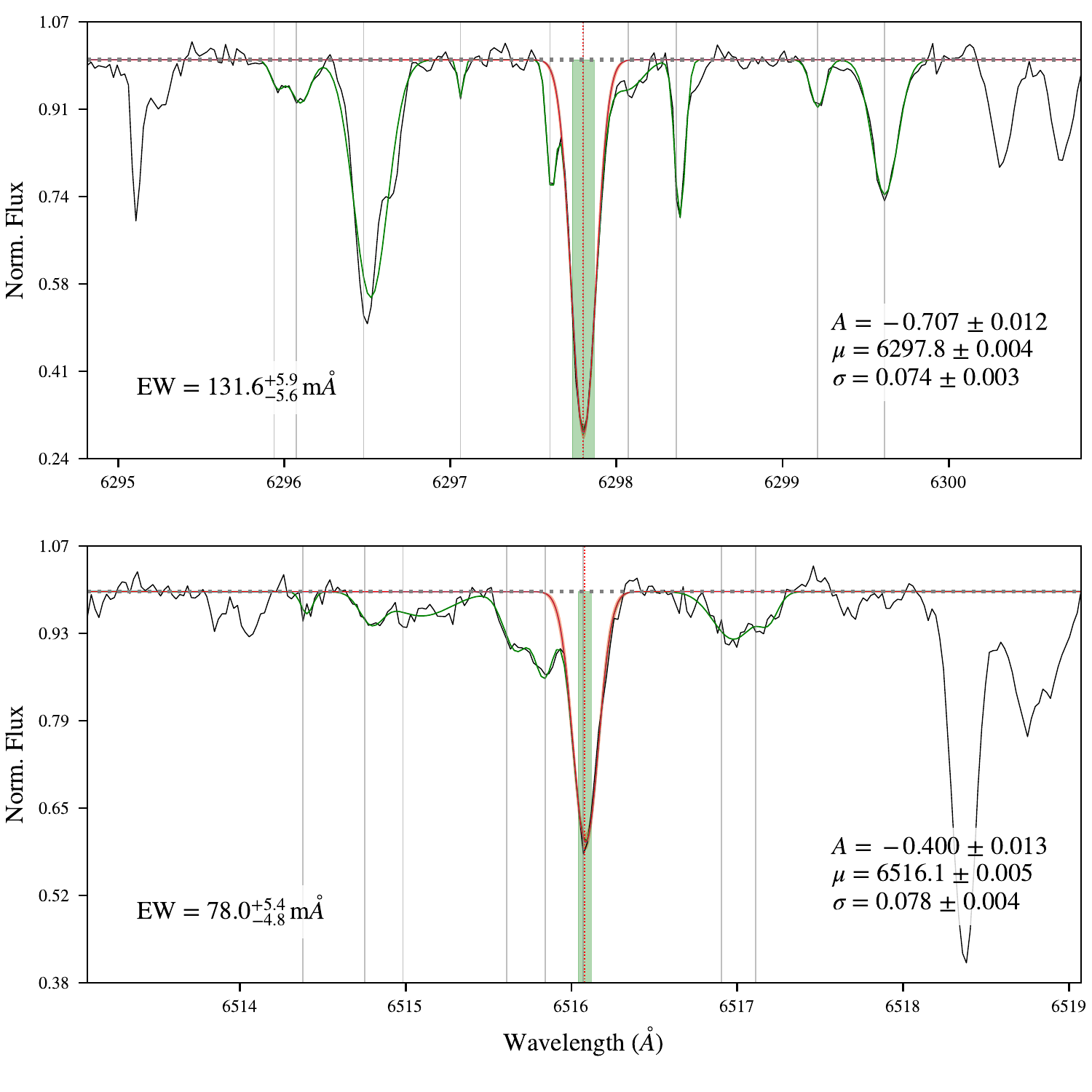}
\caption{Fit performed using the \texttt{EWComputation} module to the blended lines \FeI~ 6297.80 \angstrom ~and \FeII~ 6516.08 \angstrom, from the giant star HIP9313 taken with CHIRON.
The right-most text contains the values of $A$, $\mu$, and $\sigma$, the amplitude, mean, and standard deviation respectively of the best Gaussian fit, along with their corresponding uncertainties $e_{A}$, $e_{\mu}$, and $e_{\sigma}$.
The grey lines are the absorption lines detected in the spectral range, and the green line the global fit to the data. The red region corresponds to the fit of the line $l$ with its uncertainty, derived using the Gaussian parameters plus their uncertainties. The green block represents the equivalent width, quoted in the left-most text. The full procedure is explained in Sect.~\ref{sec:EW}.}
\label{fig:HIP9313lines}
\end{figure}

The data we used for the computation of the stellar parameters consist of the FEROS, CHIRON and HARPS spectra. In the case of FEROS data, we stacked together all the individual spectra to create a high signal-to-noise-ratio (SNR) template, which is also used to compute the RVs (see \citealt{Jones2017}), while for CHIRON and HARPS data, we simply used a high SNR observation.
The spectra were analysed using SPECIES (SJ18), which relies on the equivalent width (EW) measurement of iron lines to estimate the atmospheric parameters (\teff, \logg, \feh and \vt, the microturbulence velocity). SPECIES uses MOOG\footnote{February 2017 version} \citep{moog}, along with the equivalent widths and ATLAS9 model atmospheres \citep{ATLAS9} to solve the radiative transfer equation in local thermodynamical equilibrium (LTE) conditions. What follows is an iterative process in which the atmospheric parameters are modified until: 1) there is no correlation between the individual \FeI line abundances with the excitation potential and the reduced equivalent width (EW/$\lambda$); 2) the average abundance of \FeI and \FeII are the same, and 3) the obtained abundance of \FeI is the same as the one used to generate the model atmosphere of that iteration \citep{Gray2005}. 
\newline \indent
The EWs were measured within SPECIES, by fitting Gaussian-shaped profiles to the absorption lines using the \texttt{EWComputation} module. The method is described in Sect.~\ref{sec:EW}.
One of the challenges of studying giant stars, like the ones in this work, is line blending, which affects the line fitting and therefore the EW estimation. The \texttt{EWComputation} module has three mechanisms that help with this problem. The first one is the line detection stage, which can identify possible lines 0.1 \angstrom ~from our target line, ensuring that the possible blend is fitted separately. The second one is the tight constraints imposed on the centre of the Gaussian fit. Undetected blends will produce a shift in the centre of the fitted line, therefore the module rejects fits that are $> 0.075$ \angstrom ~away from the target line. The final one is the constraints put into the uncertainties of the line parameters. During the testing phase of the module, we saw that, in many cases, the uncertainties for the line parameters $e_{A_i}$, $e_{\mu_i}$, and $e_{\sigma_i}$ would increase for blended lines, compared to unblended-lines, therefore we imposed that all three uncertainties have to be $<0.12$ dex. In Fig.~\ref{fig:HIP9313lines} we show two cases of blended lines detected with the \texttt{EWComputation} module in the spectra of the giant star HIP9313, taken with CHIRON.
The linelist used is described in J11 and SJ18, with some modifications. We discarded the lines for which the EWs had large uncertainties ($\sigma_{\tiny {\rm EW}}/\text{EW} \ge 0.5$) for most of the stars. We were finally left with 95 \FeI and 7 \FeII lines, listed in Table~\ref{tab:linelist}. Finally, for each individual spectrum, we only used lines with $10 \leq \mbox{EW} \leq 150$ m$\angstrom$, to avoid very shallow lines that could be mistaken for noise in the continuum, and strong lines that no longer follow a Gaussian profile.  The net affect of this selection is that we are biasing against lines formed very deep, or very high in the stellar photosphere.
As a final check of the validity of our EW measurement method for giant stars, we compared our results for Arcturus with those from \citet{Ramirez2011}, using their line list and the Arcturus spectrum from \citet{Hinkle2005}. We find that our results are in agreement with those from \citet{Ramirez2011}, with a median difference of 1~m\AA, and standard deviation of 3~m\AA.
\newline \indent
The uncertainty in the atmospheric parameters derived by SPECIES is estimated from the uncertainty in the correlation between the iron abundance and the excitation potential (for the temperature), and between the abundance and the reduced EW (for the microturbulent velocity); for the metallicity, it is the spread of abundances from the \FeI lines, and for the surface gravity, it is derived from the spread of the \FeI and \FeII lines (details can be found in SJ18).
\newline \indent
The ATLAS9 model atmospheres assumes plane-parallel layers, as opposed to other models that assume a spherical geometry, like the MARCS models \citep{Gustafsson2008}. In order to check that our selection of models is not affecting the computation of parameters, we recomputed them using the MARCS models. We find that the differences between the atmospheric parameters are within their respective uncertainties, and therefore the assumption of a plane-parallel atmosphere does not have a significant effect on the EXPRESS sample of stars.

\subsection{Initial parameters}\label{sec:initial_parameters}

Before starting the estimation of the atmospheric parameters, we gather photometric and astrometric information for each target. Parallax and proper motion were obtained from Gaia DR2 \citep{GaiaDR2} or the Hipparcos mission \citep{Hipparcos}, 2MASS \textit{JHK} magnitudes from \citet{2mass}, Tycho-2 $(BV)_t$ magnitudes from \citet{tycho-2}, Str\"{o}mgren $b-y$, $m_1$, and $c_1$ from \citet{hauck1998} or \citet{Holmberg2009}, and Johnson $BV(RI)_c$ magnitudes from \citet{koen2010}, \citet{casagrande2006}, \citet{Beers2007}, or \citet{Ducati2002}.
We also correct the magnitudes for dust extinction using the maps mentioned in \citet{bovy2016}, through the \texttt{mwdust}\footnote{\url{https://github.com/jobovy/mwdust}} python package. For the case of the Gaia $G$ magnitude, which is not included within the bandpasses of \texttt{mwdust}, we used $A_G = 2.35\, E(B-V)$, from \citet{Bovy2014, bovy2016}. We preferred to compute our own extinction rather than using the $A_G$ values provided in Gaia DR2 because we find that these are overestimated for the stars in our sample (average value of $\sim 0.2$, when available), which are all nearby stars (average parallax $\sim 11$ mas), away from the galactic centre, and therefore should experience low dust extinction in average. 
Finally, if the parallaxes were obtained from Gaia DR2, we apply the systematic correction from \citet{Stassun2018}\footnote{Recently, \citet{Chan2020} computed an updated zero-point parallax correction, but the value is within the uncertainties of the parallaxes for the stars in this study, therefore we decided to keep the correction from \citet{Stassun2018} as it would not greatly affect our results.}.
\newline \indent
We use the parallax, along with the proper motion, to estimate the evolutionary state of each target following the classification scheme presented in \citet{CollierCameron2007}, which uses the photometric colour $J-H$ and the reduced proper motion (RPM), defined as $J + 5\log(\mu)$, with $J$ the apparent magnitude in the $J$-band and $\mu$ the proper motion. If the star is classified as a giant, we estimate its temperature following the relations from \citet{Alonso1999}. In Sect.~\ref{sec:results} we show how efficient that classification scheme was with our sample, and how the initial temperatures compare to the final values obtained from SPECIES.
\newline \indent
Surface gravity was inferred as the minimum value between 3.5 and Eq. 1 from SJ18. Finally, for all the stars in the giant sample, the metallicity was set to an initial value of zero. 

\subsection{Physical parameters}\label{sec:physical_parameters}

\begin{table}
\caption{Correspondance between Equivalent Evolutionary Points (EEPs) and stellar evolutionary phase.}
\label{tab:EEP}
\centering
\begin{tabular}{l l l}
\hline\hline
EEP & Phase & Abbreviation\\
\hline
1 & pre-main sequence & PMS\\
202 & zero age main sequence & ZAMS\\
353 & Intermediate age main sequence & IAMS\\
454 & terminal age main sequence & TAMS\\
605 & tip of the red giant branch & RGBTip\\
631 & zero age core helium burning & ZACHeB\\
707 & terminal age core helium burning & TACHeB\\
\hline
\end{tabular}
\end{table}

The physical parameters, namely mass, radius, age, luminosity, and evolutionary stage, were estimated using the latest version of the \texttt{isochrones}\footnote{\url{https://github.com/timothydmorton/isochrones}} package \citep{Morton2015}. This package creates a model of the star based on the atmospheric parameters, and estimates its physical parameters by interpolating through a grid of MESA Isochrones and Stellar Tracks \citep[MIST,][]{Dotter2016} using MultiNest \citep{feroz2009, buchner2014}. The final values for each parameter will be drawn from the obtained distributions.
The newest version of this package incorporates the interpolation of the isochrone grids using the equivalent evolutionary points (EEPs), which indicates the evolutionary stage of the star.

As described in \citet{Dotter2016}, EEPs are points in stellar evolution that can be identified in different evolutionary tracks, and are created to allow for the interpolation between tracks defined for different initial stellar masses. In the case of the MIST models, they are divided into primary EEPs, defined following a physical motivation, and secondary EEPs, that provide a uniform spacing between the primary EEPs.
The phases that are relevant for this work are:
\begin{itemize}
    \item Terminal Age Main Sequence (TAMS), defined as the point where the central H mass fraction is $10^{-2}$, after which the star begins the burning of H outside the core and enters the RGB phase.
    \item Tip of the Red Giant Branch (RGBTip), the point at which the stellar luminosity reaches a maximum (or \teff reaches a minimum), but before sustained He burning in the core has started. For low-mass stars ($M < 2.00 \, M_{\odot}$, \citealt{Paxton2010}), this point denotes the beginning of the helium flash, when He burning begins under degenerate conditions in the core.
    \item Zero Age Core Helium Burning (ZACHeB), where sustained core He burning begins, which will be referred to as the HB phase.
    \item Terminal Age Core Helium Burning (TACHeB), when the central He fraction is $10^{-4}$, which denotes the end of the core helium burning. Subsequent evolutionary stages include the burning of He outside the core, and carbon burning for high-mass stars. 
\end{itemize}
A more detailed description and processes involved in defining the EEPs can be found in \citet{Dotter2016}.
The use of EEPs in \texttt{isochrones} allows for the correct interpolation in the isochrone grids for the evolved stellar phases\footnote{\url{https://isochrones.readthedocs.io/en/latest/modelgrids.html}}.
As \citet{Dotter2016} pointed out, inaccuracy in the interpolation happens because of the wide scale of changes the star goes through after the main sequence, all at very short timescales, compared to the time spent in the main sequence. During the post-main sequence phases, tracks for very similar initial masses will lie very close to each other but, for a certain age, will show stars in completely different evolutionary stages. 
A simple interpolation scheme that does not take into account the faster evolutionary phase after the MS might not be able to fully reproduce the morphology of the evolution track. This is one of the reasons for the discrepancy between the results presented here and those in J11.
The numbers referred as EEPs in the rest of the text correspond to the secondary EEPs.

During the interpolation process, we adopted a prior mass distribution following a Salpeter initial mass function (IMF), with $\alpha = 2.35$. Similarly, we adopted a prior metallicity distribution, based on the local disk, following \citet{Casagrande2011}\footnote{For a given metallicity value $x$, the probability density function takes the form
$P(x) = w_0(w_1 N_1(x) + w_2 N_2(x))$, where $N_i = G(\mu_i, \sigma_i)$ represents a Gaussian distribution with parameters ($\mu_i, \sigma_i$), and the $w_i$ terms are normalisation factors.
}
For the stellar age we adopted a flat prior in logarithmic scale, with bounds given as ($10^5$ - $10^{10.15}$) yrs. Finally, the EEP has an uniform probability distribution, with 353 as the lower bound for giant stars (otherwise 200, to ensure no pre-main sequence stages), and 1710 as the upper bound. 
The atmospheric parameters derived previously (\teff, \logg, and \feh), along with the photometric magnitudes and parallax (Sect.~\ref{sec:initial_parameters}), were added to the likelihood function as $\ln L \propto -\frac{1}{2}\sum_i (x_i - I_i)^2/\sigma_i^2$, where ($x_i$, $\sigma_i$) correspond to the observed and computed properties, and $I_i$ is the model prediction for the $i$'th parameter \citep{Montet2015}.
\newline \indent
One of the outputs from the physical parameter estimation is the EEP distribution for each star.
Table~\ref{tab:EEP} shows the correspondence between EEP and evolutionary phase. For the rest of the analysis in the paper, we consider 454 $<$ EEP $<$ 631 to be the RGB phase, and 631 $\leq$ EEP $\leq$ 707 to be the HB phase. We also infer the probability of a star to be either in the RGB or the HB phase as $p_{\text{state}} = \sum \text{EEP}_{\text{state}}/ \sum N$, where $N$ is the total EEP distribution size distribution. 

\subsection{Rotational and macroturbulent velocity} 
We followed the method described in SJ18 to compute the rotational and macroturbulent velocities, which we summarise here. 
The macroturbulent velocity is computed following Eq. 1 in \citet{dosSantos2016}, and for the cases when we obtain a very small or negative value, indicating that the temperature and/or surface gravity is outside of the range of applicability, we adopt Eq. 2 from \citet{Brewer2016}. The rotational velocity was estimated by fitting rotationally broaden synthetic profiles to five different absorption lines. 
To obtain an estimate of the uncertainty in the rotational velocities, we first vary the normalized flux of each line by an amount $dy$, with $dy$ being drawn from a normal distribution with zero mean and width given by 1/SNR, with SNR the signal-to-noise ratio of the spectrum. We then recompute the rotational velocity a total of 1000 iterations, obtaining a distribution of rotational velocities. The uncertainty is then taken as the standard deviation of the distribution. The final rotational velocity will correspond to the average of the velocities obtained for the five lines, with their corresponding errors. More details of the fitting procedure can be found in SJ18.

\section{Results}\label{sec:results}

   \begin{figure*}
   \centering
   \includegraphics[width=\hsize]{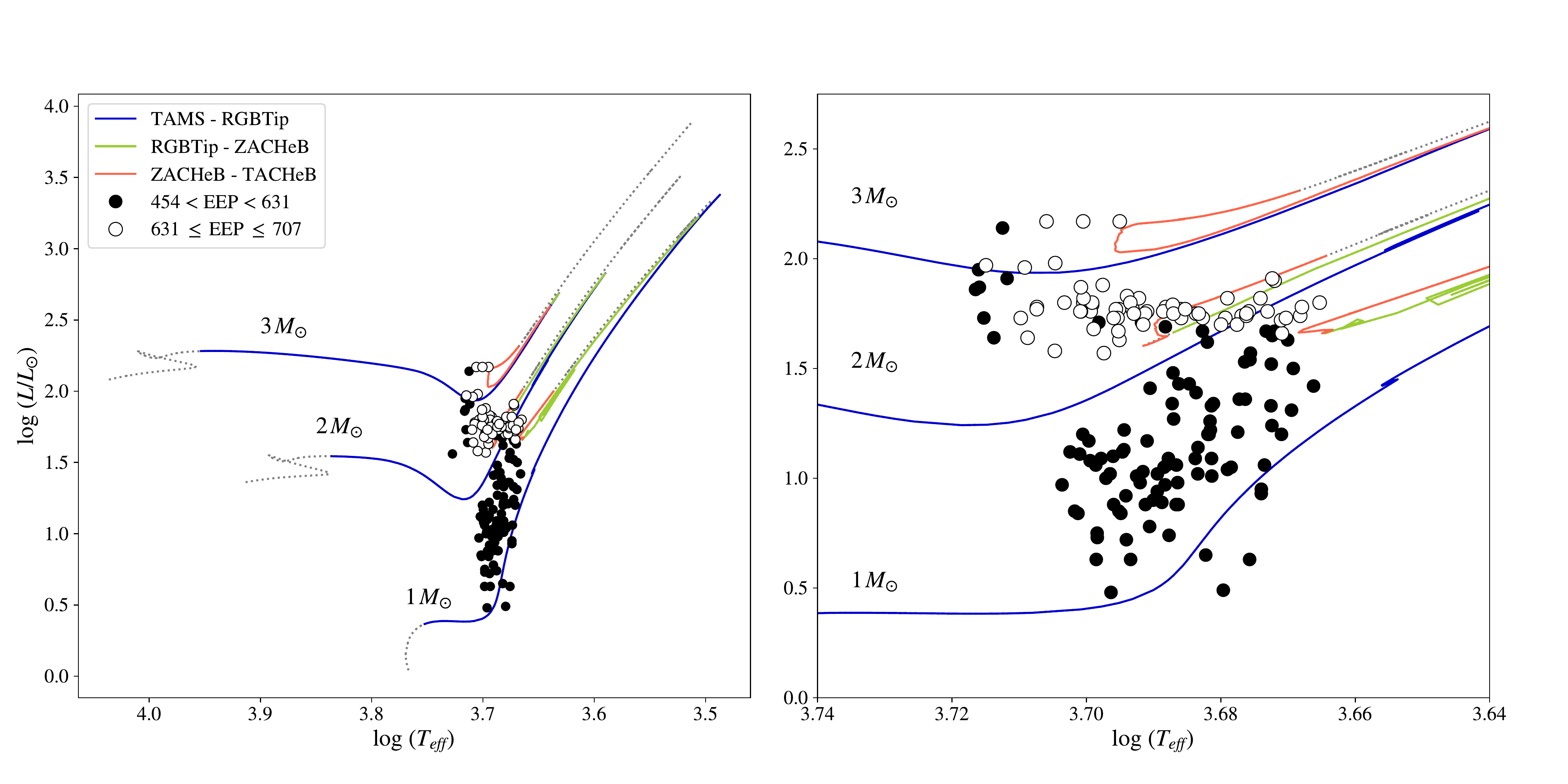}
      \caption{H-R diagram of the giant star sample. The right panel is a close-up of the area where the stars in the sample are located. Open circles represent RGB stars, and filled circles HB stars. The tracks plotted have [Fe/H] = 0.0, and the different colours represent different phases in the stellar evolution. Grey lines represent the stages before the main sequence evolution (up to the TAMS, Table~\ref{tab:EEP}), and right after the end of core helium burning (TACHeB). For the meaning of the abbreviations and EEP number, see Table
     ~\ref{tab:EEP}.}
         \label{fig:HRdiagram}
   \end{figure*}

\begin{table}
\caption{Values obtained for the Sun spectra using SPECIES. Only average values are shown. Individual results for each spectrum are shown on Table~\ref{tab:sunall}}
\label{tab:sun_average}
\centering
\begin{tabular}{l l c}
\hline\hline
Parameter & Unit & Value \\
\hline
Temperature & K & 5770 $\pm$ 35\\
{[Fe/H]} & dex & -0.03 $\pm$ 0.04\\
logg & cm\,s$^{-2}$ & 4.36 $\pm$ 0.04\\
$\xi_t$ & km s$^{-1}$ & 0.82 $\pm$ 0.05\\
v$\,\sin\,$i & km s$^{-1}$ & 3.61 $\pm$ 0.19\\
$v_{\text{mac}}$ & km s$^{-1}$ & 3.19 $\pm$ 0.15\\
Mass & $M_{\odot}$ & 0.97 $\pm$ 0.02 \\
Radius & $R_{\odot}$ & 1.06 $\pm$ 0.05\\
Age & Gyr & 6.9 $\pm$ 1.5\\
log L & $L_{\odot}$ & 0.05 $\pm$ 0.06\\
\hline
\end{tabular}
\end{table}

The sample of giant stars from the EXPRESS project was first described and analysed in J11, using a broadly similar procedure to here for the estimation of atmospheric and physical parameters. Here we improve on this analysis by using updated versions of the atmospheric models and the MOOG code, unavailable when J11 was published. We also estimate the equivalent widths using the method described in Sect. \ref{sec:EW}, instead of using ARES \citep{Sousa2007}, as was done in J11.
Additionally, here we use an updated version of the line list. We removed 68 lines from the original list used in J11, 54 \FeI and 14 \FeII lines, due to the consistently large uncertainties in the measured equivalent widths ($\sigma_{\text{EW}}/\text{EW} > 0.5$). The updated line list is given in Table~\ref{tab:linelist}.
In addition, here we used higher quality spectra to measure the EWs, in terms of resolving power and SNR. For $\sim$ 30 stars presented in J11 we used FECH low SNR spectra at a resolution lower than FEROS and HARPS. As a consequence, the EWs and thus the derived atmospheric and physical parameters are largely uncertain. As an example we highlight the case of the planet-host star HIP\,105854, for which we used a low quality FECH spectrum for the analysis. In J11 the obtained mass was 2.1$\pm$0.1 \msun, a substantial overestimate of the mass obtained here (0.97$^{+0.09}_{-0.04}$ \msun). This particular case was also noted by \citet{Campante2019}, who used asteroseismology to derive a mass of 1.00$\pm$ 0.16 \msun, if the star is in HB, as obtained here ($p_{\rm HB} = 0.88$; see Table \ref{tab:expressfull}). 
Finally, in the case of the physical parameters (like mass and radius), here we use the Gaia DR2 parallaxes, newer extinction maps and we also employ a different set of evolutionary models.

The HR diagram for the EXPRESS stars is shown in Fig.~\ref{fig:HRdiagram}. A sample of the catalogue produced by SPECIES, with a subset of columns, is shown in Table~\ref{tab:expressfull}. We also include in Table~\ref{tab:sun_average} the average values obtained for a set of Sun spectra, taken using different methods and instruments. The individual results are listed in Table~\ref{tab:sunall}.

\subsection{Comparison with the literature}

We first analyse the stars from EXPRESS with results from interferometry and asteroseismology, which we use as a benchmark for the results obtained using our techniques.  
To this comparison we have also added a few other stars with spectra available from FEROS and HARPS, which are not necessary evolved stars.
The list of stars with the results from SPECIES (for interferometry and asteroseismology) are listed in Table~\ref{tab:test_SPECIES}, and the values from the literature in Table~\ref{tab:test_literature}.
We then compare our results with other spectroscopic studies.

\subsubsection{Interferometry}

   \begin{figure}
   \centering
            \includegraphics[width=\hsize]{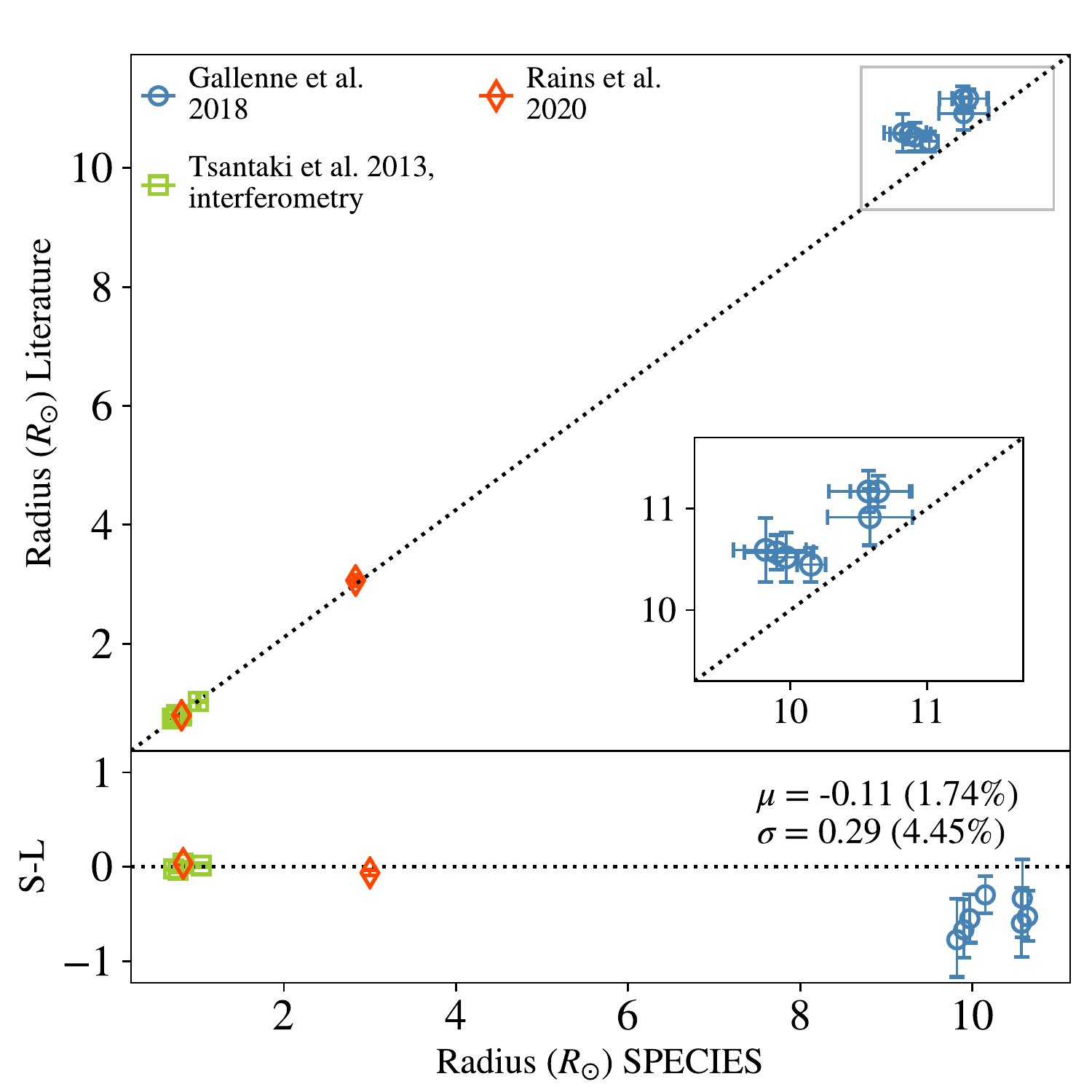}
      \caption{Top panel: Comparison between the radius measurement from SPECIES and from works using interferometry, for the stars in our test sample.
               Black dotted line is the 1:1 relation.
               The inset correspond to a zoom of the squared regions.
               Bottom panel: Difference SPECIES-literature. Uncertainties are computed considering both SPECIES and literature values ($\sqrt{\sigma_{\text{SPECIES}}^2 + \sigma_{\text{Lit}}^2}$).
               The quoted text is the mean and standard deviation of the residuals, and their percentage from the mean radius value.
              }
         \label{fig:interferometry}
   \end{figure}

\begin{table}
\caption{Offsets between the radius from SPECIES, and from interferometry studies. $\mu$, $\sigma_{\mu}$, and $\sigma$ are the mean, error in the mean, and standard deviation of the distribution from the residuals SPECIES-Literature, respectively.}
\label{tab:interferometry}
\centering
\begin{tabular}{l | cccc}
\hline\hline
Catalogues & \multicolumn{3}{c}{Radius ($R_{\odot}$)} & N\\
 & $\mu$ & $\sigma_{\mu}$ & $\sigma$ & \\
 \hline
\citet{Gallenne2018} & -0.54 & 0.13 & 0.16 & 7 \\
\citet{Rains2020} & -0.02 & 0.02 & 0.05 & 2 \\
\citet{Tsantaki2013} & -0.00 & 0.01 & 0.03 & 4 \\
Total\tablefootmark{a} & -0.11 & 0.01 & 0.29 & 12 \\
\hline
\end{tabular}
\tablefoot{
\tablefoottext{a}{The mean and error for the total values were computed using weights given by $w_i=(x_i/\sigma_i)^2$, with ($x_i$, $\sigma_i$) the value and uncertainty of each point $i$.}}
\end{table}

Precise stellar angular sizes can be obtained through long baseline interferometry, which can then be combined with parallax data and broadband photometry to estimate the size of a star. 
We compare a total of 12 targets (5 from the EXPRESS sample and 7 new targets with data from FEROS) with the radius from \citet{Gallenne2018}, \citet{Tsantaki2013}, and \citet{Rains2020}, and the results are shown in Fig.~\ref{fig:interferometry} and Table~\ref{tab:interferometry}.
 In the case of \citet{Tsantaki2013}, the authors do not explicitly list the radius found through interferometry, but show the limb-darkened angular diameter $\theta_{\text{LD}}$ used, taken from the literature \citep[the corresponding references are found in][]{Tsantaki2013}. We combined $\theta_{\text{LD}}$ with parallax information from either Gaia or Hipparcos (whichever was the most precise) to obtain the stellar radius.

Overall, for interferometry we find a difference in the radius between SPECIES and the literature of $\Delta\, R_{\text{INTER}} = -0.11 \pm 0.29\,R_{\odot}$, below the 2\% level from the mean radius values.
Here, the values are quoted as $\mu \pm \sigma$, the mean and standard deviation of the residual distribution, respectively. The comparison with interferometry is shown in Fig.~\ref{fig:interferometry}.
We find that the differences in radius are within a standard deviation of the individual values (Table~\ref{tab:interferometry}), where our results are 0.54 $R_{\odot}$ ($\sim $5\%) smaller \citet{Gallenne2018}.
These substantial differences are not seen when comparing with \citet{Tsantaki2013} and \citet{Rains2020}, but those targets correspond to main sequence stars, whereas the targets from \citet{Gallenne2018} are giant stars.
The source of our discrepancy with \citet{Gallenne2018} is not yet clear, but could be related to the calibration or limb-darkening model used for the estimation of the angular sizes.

\subsubsection{Asteroseismology}
   
   \begin{figure*}
   \centering
            \includegraphics[width=\textwidth]{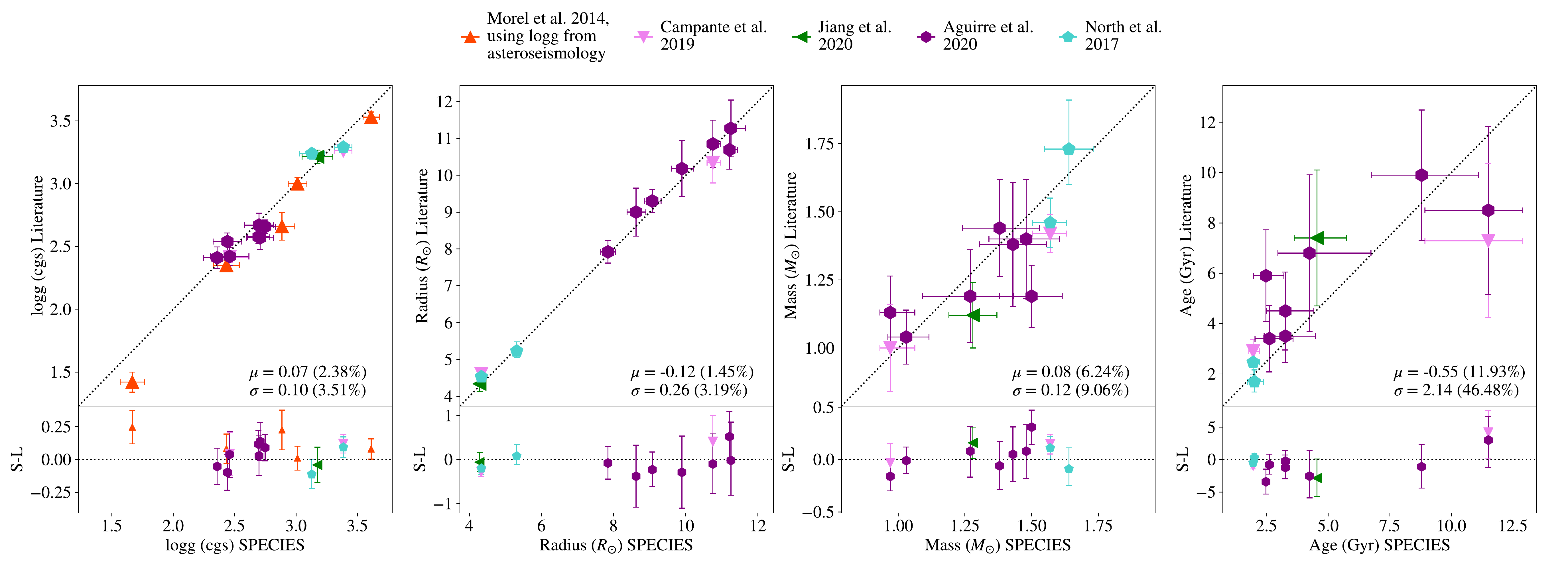}
      \caption{Top panels: Comparison between SPECIES and the asteroseismology studies, for the stars in our test sample.
               Black dotted line in each panel is the 1:1 relation.
               Bottom panels: Difference SPECIES-literature, same as in Fig.~\ref{fig:interferometry}.
               The quoted text are the mean and standard deviation of the residuals for each quantity, and their percentage from the mean values.
               Surface gravity values were not given in \citet{Aguirre2020} and \citet{North2017}, but were computed from their masses and radius.
              }
         \label{fig:asteroseismology}
   \end{figure*}

\begin{table*}
\caption{Offsets between the results from SPECIES, and from asteroseismic studies. $\mu$, $\sigma_{\mu}$, and $\sigma$ are the mean, error in the mean, and standard deviation of the distribution from the residuals SPECIES-Literature, respectively.}
\label{tab:asteroseismology}
\centering
\begin{tabular}{l|ccc|ccc|ccc|ccc|c}
\hline\hline
Catalogues & \multicolumn{3}{c}{logg} & \multicolumn{3}{c}{Radius} & \multicolumn{3}{c}{Mass} & \multicolumn{3}{c}{Age} & N \\
& \multicolumn{3}{c}{(cgs)} & \multicolumn{3}{c}{($R_{\odot}$)} & \multicolumn{3}{c}{($M_{\odot}$)} & \multicolumn{3}{c}{(Age)} & \\
 & $\mu$ & $\sigma_{\mu}$ & $\sigma$ & $\mu$ & $\sigma_{\mu}$ & $\sigma$ & $\mu$ & $\sigma_{\mu}$ & $\sigma$ & $\mu$ & $\sigma_{\mu}$ & $\sigma$ & \\
 \hline
\citet{Morel2014} & 0.13 & 0.05 & 0.09 & & & & & & & & & & 5 \\
\citet{Campante2019} & 0.08 & 0.09 & 0.04 & 0.06 & 0.30 & 0.35 & 0.06 & 0.10 & 0.09 & 1.63 & 2.02 & 2.58 & 2 \\
\citet{Jiang2020} & -0.04 & & & -0.06 & & & 0.16 & & & -2.86 & & & 1 \\
\citet{Aguirre2020} & 0.04 & 0.05 & 0.08 & -0.08 & 0.25 & 0.27 & 0.04 & 0.08 & 0.14 & -0.91 & 1.09 & 1.89 & 7 \\
\citet{North2017} & -0.01 & 0.07 & 0.10 & -0.06 & 0.15 & 0.14 & 0.01 & 0.12 & 0.10 & -0.10 & 0.47 & 0.41 & 2 \\
Total\tablefootmark{a} & 0.07 & 0.01 & 0.10 & -0.12 & 0.01 & 0.26 & 0.08 & 0.03 & 0.12 & -0.55 & 0.11 & 2.14 & 15 \\
\hline
\end{tabular}
\tablefoot{
\tablefoottext{a}{The mean and error for the total values were computed using weights given by $w_i=(x_i/\sigma_i)^2$, with ($x_i$, $\sigma_i$) the value and uncertainty of each point $i$.}}
\end{table*}

The asteroseismology technique consists of looking at stellar oscillation spectra from either spectroscopy or photometry, and measuring two global oscillation quantities: the average frequency separation ($\Delta\nu$), and the frequency corresponding to the maximum oscillation power ($\nu_{\text{max}}$). In the case of solar-like oscillations, these values can then be used to estimate the stellar density and surface gravity by using scaling relations \citep{Ulrich1986, Brown1991, Kjeldsen1995}, and by combining both quantities it is possible to obtain the stellar mass and radius, in a nearly model-independent approach.
Another method widely used in the literature to obtain the stellar parameters is by using grid-based modelling techniques, that take as an input the individual frequency determinations or the global asteroseismic parameters, plus the effective temperature and metallicity, and other observational constraints. The scaling relations can also be included in the grid-based modelling.
Even though this method is not model-independent, it has been found that the results given by the use of different pipelines, all using different stellar evolution models, are all in good agreement \citep{Pinsonneault2018, SilvaAguirre2015}.
When the analysis is made using the grid-modelling approach, it is possible to also determine the stellar age. Age is a difficult parameter to fit for \citep{Soderblom2010}, but asteroseismology has been so far the most precise method for its estimation, with uncertainties in the order of $\sim 25$\% \citep{Silva2016}. 
The other properties derived from asteroseismology (like mass and radius) have been shown to be accurate to the $\sim$ 10\% level \citep{Pinsonneault2018}.
A more detailed discussion on the subject of asteroseismology and the different pipelines available for the parameter estimation can be found in \citet{Chaplin2013, SilvaAguirre2020}, and references therein.
\newline \indent
We compared our estimates for the mass, radius, logg, and age for 15 targets (9 from EXPRESS, and 6 new targets with data taken with FEROS) with values computed using asteroseismology from \citet{Morel2014}, \citet{Campante2019}, \citet{Jiang2020}, \citet{Aguirre2020}, and \citet{North2017}. 
In \citet{Morel2014}, the surface gravity is computed from the scaling relation that uses $\nu_{\text{max}}$ and the stellar effective temperature. The temperature is computed from high-resolution spectroscopy, following the same method as in this work. As the authors do not list the values for $\Delta\nu$ and $\nu_{\text{max}}$, it was not possible for us to separate the stellar mass and radius from the surface gravity. 
In \citet{Campante2019}, \citet{Jiang2020}, \citet{Aguirre2020}, and \citet{North2017}, the authors used temperature and metallicity values from the literature, and were used, together with the global asteroseismic parameters and astrometry information from Gaia (Hipparcos in the case of \citealt{North2017}), to retrieve the stellar parameters using grid-based modelling techniques. For \citet{Aguirre2020} and \citet{North2017}, the logg values used for the comparison were estimated from their mass and radius results.
\newline \indent
Figure~\ref{fig:asteroseismology} shows the comparison between SPECIES and the asteroseismic literature, and Table~\ref{tab:asteroseismology} lists the average differences.
We find $\Delta\,\log\text{g}_{\text{ASTERO}} = 0.07 \pm 0.10$~cgs, $\Delta\, R_{\text{ASTERO}} = -0.12 \pm 0.26\,R_{\odot}$, $\Delta\, M_{\text{ASTERO}} = 0.08 \pm 0.12 \,M_{\odot}$, and $\Delta\, \text{Age}_{\text{ASTERO}} = -0.55 \pm 2.14$~Gyr.
These offsets show agreement at the 2.4\% level for the surface gravity, 1.5\% for the radius, 6.2\% for the mass, and 11.9\% for the age. The standard deviations are within the 5\% level for the surface gravity and the radius, within the 10\% for the mass, and just within the 47\% level for the age.
We also highlight that the average offset with the radius agrees with the one from interferometry.
For all the parameters, we are within a standard deviation of the residuals (SPECIES-Literature), which leads us to conclude that our results agree with the asteroseismology ones. 
\newline \indent
As we mentioned in the introduction, it is especially important to check the mass measurements of giant stars, computed using stellar evolution grids, with results from other methods. We find that our mass estimates, as a whole, agree with the asteroseismic values within the uncertainties (with the exception of HIP4293, from \citealt{Aguirre2020}). This result validates the mass estimates with SPECIES, which are shown for the whole EXPRESS sample in the next section.

\subsubsection{Spectroscopy}

   \begin{figure*}
   \centering
            \includegraphics[width=\textwidth]{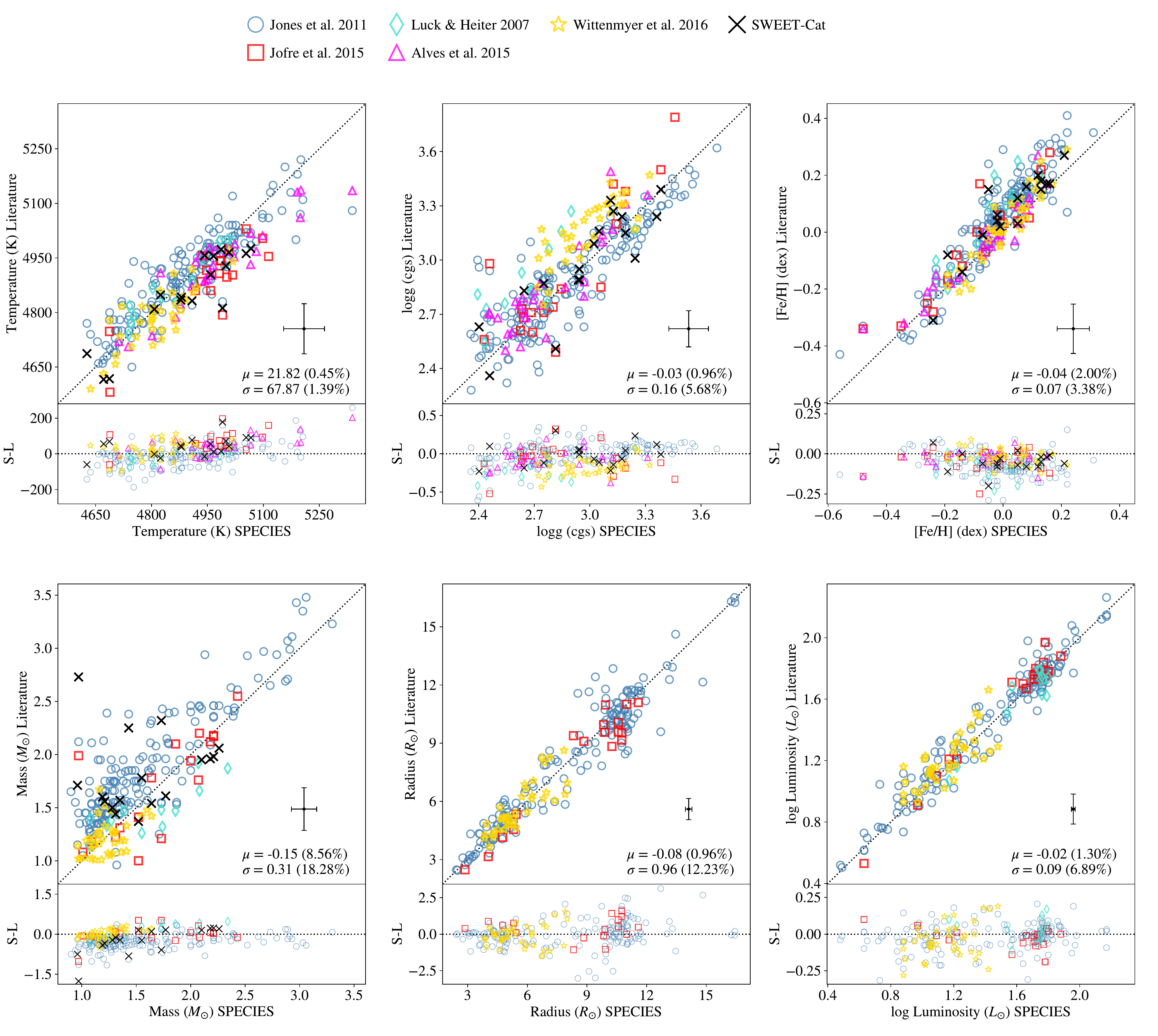}
      \caption{Comparison between SPECIES and the literature for the EXPRESS stars. Bottom panels show the difference SPECIES-Literature. The errorbar shows the average uncertainties in the points.
      The quoted text are the mean and standard deviation of the residuals for each quantity, and their percentage from the mean values.
              }
         \label{fig:giant_stars}
   \end{figure*}

\begin{table*}
\caption{Residuals between our results and the literature. $\mu$, $\sigma_{\mu}$, and $\sigma$ correspond to the mean, error in the mean, and standard deviation of the residual distribution, respectively. The last column shows the number of stars in common with each catalogue.
}
\label{tab:offsets}
\centering
\resizebox{\textwidth}{!}{
\begin{tabular}{l|rrr|rrr|rrr|rrr|rrr|rrr|c}
\hline\hline
Cata- & \multicolumn{3}{c}{Temperature} & \multicolumn{3}{c}{logg} & \multicolumn{3}{c}{[Fe/H]} & \multicolumn{3}{c}{$M/M_{\odot}$} & \multicolumn{3}{c}{$R/R_{\odot}$} & \multicolumn{3}{c}{log $L/L_{\odot}$} & N\\
logue\tablefootmark{a} & \multicolumn{3}{c}{(K)} & \multicolumn{3}{c}{(cgs)} & \multicolumn{3}{c}{(dex)} & & & &\\
 & $\mu$ & $\sigma_{\mu}$ & $\sigma$ & $\mu$ & $\sigma_{\mu}$ & $\sigma$ & $\mu$ & $\sigma_{\mu}$ & $\sigma$ & $\mu$ & $\sigma_{\mu}$ & $\sigma$ & $\mu$ & $\sigma_{\mu}$ & $\sigma$ & $\mu$ & $\sigma_{\mu}$ & $\sigma$ & \\
\hline
J11 & -4.09 & 7.56 & 69.82 & 0.00 & 0.01 & 0.16 & -0.06 & 0.01 & 0.07 & -0.28 & 0.01 & 0.27 & -0.02 & 0.02 & 1.03 & -0.021 & 0.001 & 0.090 & 166 \\
J15 & 67.70 & 14.77 & 60.58 & -0.06 & 0.03 & 0.20 & -0.05 & 0.02 & 0.07 & -0.00 & 0.05 & 0.33 & 0.40 & 0.15 & 0.73 & -0.026 & 0.025 & 0.068 & 17 \\
L07\tablefootmark{b} & -13.32 & 14.83 & 33.03 & -0.21 & 0.03 & 0.14 & -0.10 & 0.03 & 0.06 & 0.14 & 0.05 & 0.26 & & & & 0.035 & 0.004 & 0.075 & 12 \\
A15\tablefootmark{c} & 49.75 & 12.38 & 56.71 & -0.07 & 0.03 & 0.12 & -0.02 & 0.01 & 0.05 & & & & & & & & & & 27 \\
W16 & 27.10 & 9.54 & 47.84 & -0.18 & 0.02 & 0.10 & -0.02 & 0.01 & 0.05 & 0.01 & 0.02 & 0.18 & -0.10 & 0.02 & 0.69 & -0.044 & 0.002 & 0.113 & 37 \\
SC\tablefootmark{d} & 43.47 & 20.25 & 53.23 & -0.01 & 0.05 & 0.15 & -0.05 & 0.02 & 0.06 & -0.25 & 0.07 & 0.50 & & & & & & & 17 \\
Total\tablefootmark{e} & 21.82 & 0.01 & 67.87 & -0.03 & 0.01 & 0.16 & -0.04 & 0.01 & 0.07 & -0.15 & 0.01 & 0.31 & -0.08 & 0.01 & 0.96 & -0.018 & 0.001 & 0.093 &  \\
\hline
\end{tabular}}
\tablefoot{
\tablefoottext{a}{J11: \citet{Jones2011}, J15: \citet{Jofre2015}, L07: \citet{Luck2007}, A15: \citet{Alves2015}, W16: \citet{Wittenmyer2016}, SC: SWEET-Cat}.\\
\tablefoottext{b}{The parameters used correspond to the spectroscopic ones, and the mass obtained using the spectroscopic \teff.}\\
\tablefoottext{c}{Only the results from using the line-list from \citet{Tsantaki2013}.}\\
\tablefoottext{d}{We only considered the data from their homogeneous sample, meaning that the parameters were derived following the uniform methodology described in \citet{Santos2013}. The individual stars were analysed in \citet{Mortier2013a}, \citet{Andreasen2017}, and \citet{Sousa2018}.}\\
\tablefoottext{e}{The mean and error for the total values were computed using weights given by $w_i=(x_i/\sigma_i)^2$, with ($x_i$, $\sigma_i$) the value and uncertainty of each point $i$.}}
\end{table*}

We compared our results for the EXPRESS sample with works in the literature that use spectroscopy to derive the stellar parameters. These are \citet{Jones2011}, \citet{Jofre2015}, \citet{Luck2007}, \citet{Alves2015}, \citet{Wittenmyer2016}, and the SWEET-Cat\footnote{\url{https://www.astro.up.pt/resources/sweet-cat/}} catalogue \citep{Santos2013}, a compilation of stellar parameters for stars with planets discovered in the literature.
In the case of \citet{Luck2007}, we compared with the parameters derived using spectroscopy, and from \citet{Alves2015}, we used the results from the \citet{Tsantaki2013} line list, as those are the ones adopted by the authors.
Figure~\ref{fig:giant_stars} shows the comparison with our results and the mentioned works, for six of the parameters computed by SPECIES.

All these works from the literature use the same method for the derivation of the atmospheric parameters as in this paper, and differences arise in the source of the data (different instruments and S/N), the measurement of the equivalent widths, the model atmospheres, and the iron line list used. In the case of the physical parameters, when available, they were estimated using different stellar evolution models.
The average differences, with their uncertainties, and dispersion in the residuals (SPECIES - Literature) are listed in Table~\ref{tab:offsets}.

For the temperature, we find that our results are in general larger than the literature ($\Delta T_{\text{eff}} = 22 \pm 68$~K), but the difference is within a standard deviation of the data. In the case of the surface gravity and metallicity, our estimates are smaller than the literature, but both are within $1\sigma$ from zero, and represent less than 5\% of the average logg and metallicity values for our sample. For the mass, our results are smaller than the literature ($\Delta M = -0.15 \pm 0.31 \,M_{\odot}$), which is mostly driven by the comparison with J11, which we would comment more on the next paragraph. The difference of $-0.25\,M_{\odot}$ with respect to SWEET-Cat is driven mostly by four stars, two of which have large uncertainties ($\sigma \geq 0.45\, M_{\odot}$), and if we remove those two objects the difference is reduced to $-0.11\,M_{\odot}$. Finally, for both the radius and luminosity we find that our results agree with the literature values at the 2\% level. 

As we mentioned earlier, there has been some debate in the literature as to whether the masses of giant stars are overestimated when using stellar evolution codes, partly due to the challenge that poses interpolating through the post-main sequence tracks. This can be seen in our comparison with J11. We find an overall mass difference of $\Delta M_{\text{J11}} = -0.28 \pm 0.27\,M_{\odot}$, with some targets being over 1 $M_{\odot}$ smaller than in J11.
This mass difference could be linked to what we already mentioned in the introduction: stellar evolution tracks are degenerate for giant stars. This means that tracks corresponding to different masses occupy the same region in the HR diagram. Although incorporating precise values for the temperature, surface gravity, and metallicity to the interpolation helps in achieving a more accurate mass determination \citep{Sousa2015}, this might not be enough.
As we mentioned in Sect.~\ref{sec:physical_parameters}, the sampling of the evolution tracks in terms of the evolutionary state (the EEP in our case) helps in dealing with some of the complexity, as it takes into account, for example, how brief the post main sequence evolution is, compared to the total stellar lifetime \citep{Girardi2013, Dotter2016}. 

\subsection{Comparison between logg and logg$_{\text{tri}}$}

\begin{figure}
    \centering
    \includegraphics[width=\hsize]{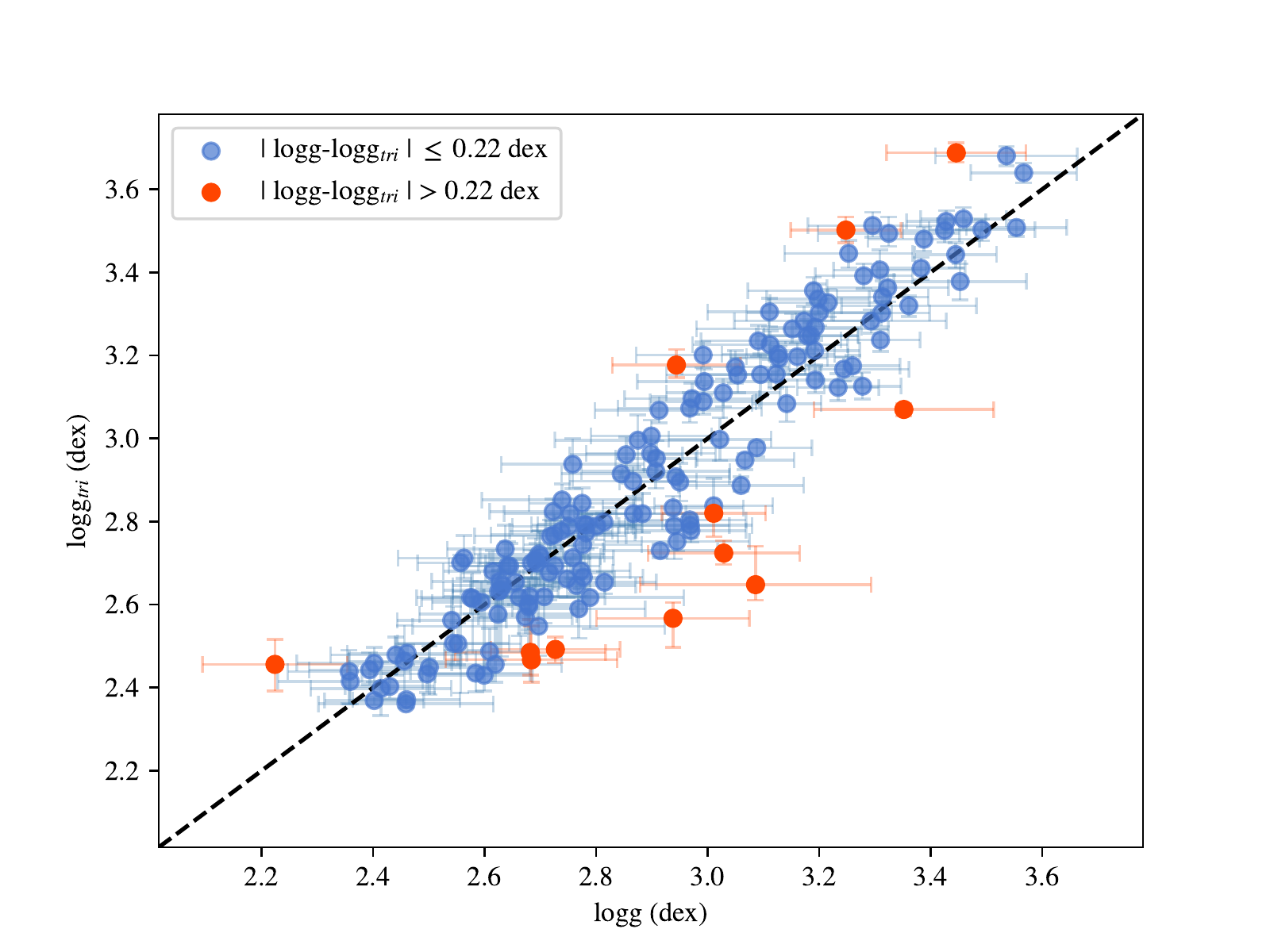}
    \caption{Comparison between \logg and \loggtri. The black dashed line represents the 1:1 relation.}
    \label{fig:comparison_logg}
\end{figure}

Previous works have shown a systematic difference between the surface gravity derived from the ionization balance of iron lines (spectroscopic \logg), and from 
interpolation in a grid of evolutionary models, usually called trigonometric \logg (e.g. \citealt{Gratton1996, Sestito2006, Jones2011, Jofre2014, Tsantaki2019, Slumstrup2019}). Some of the proposed explanations for this disagreement are the line list used to compute the \FeII abundance, or problems with the fitting of the individual absorption lines, like line blending or incorrect placement of the continuum \citep{Jones2011,Tsantaki2019}, among others.
Here, we also compared our derived spectroscopic and trigonometric logg's.
Figure~\ref{fig:comparison_logg} shows our comparison between these two independent values. As can be seen we find a good agreement between them, with an average difference of only 0.001 cgs, well below the individual mean 1-$\sigma$ uncertainty. We find a few exceptions, for which $|$\logg-\loggtri$| > 0.22$ cgs. This value was derived in SJ18 after studying the difference in both quantities. For those stars, \loggtri was adopted and the stellar parameters were recomputed, following SJ18. We found that by imposing \loggtri as the final \logg value, the rest of the parameters (\teff, \feh, and the other physical parameters) were barely affected, with changes that are within their respective uncertainties.

\subsection{Initial parameters effectiveness}

   \begin{figure}
   \centering
   \includegraphics[width=\hsize]{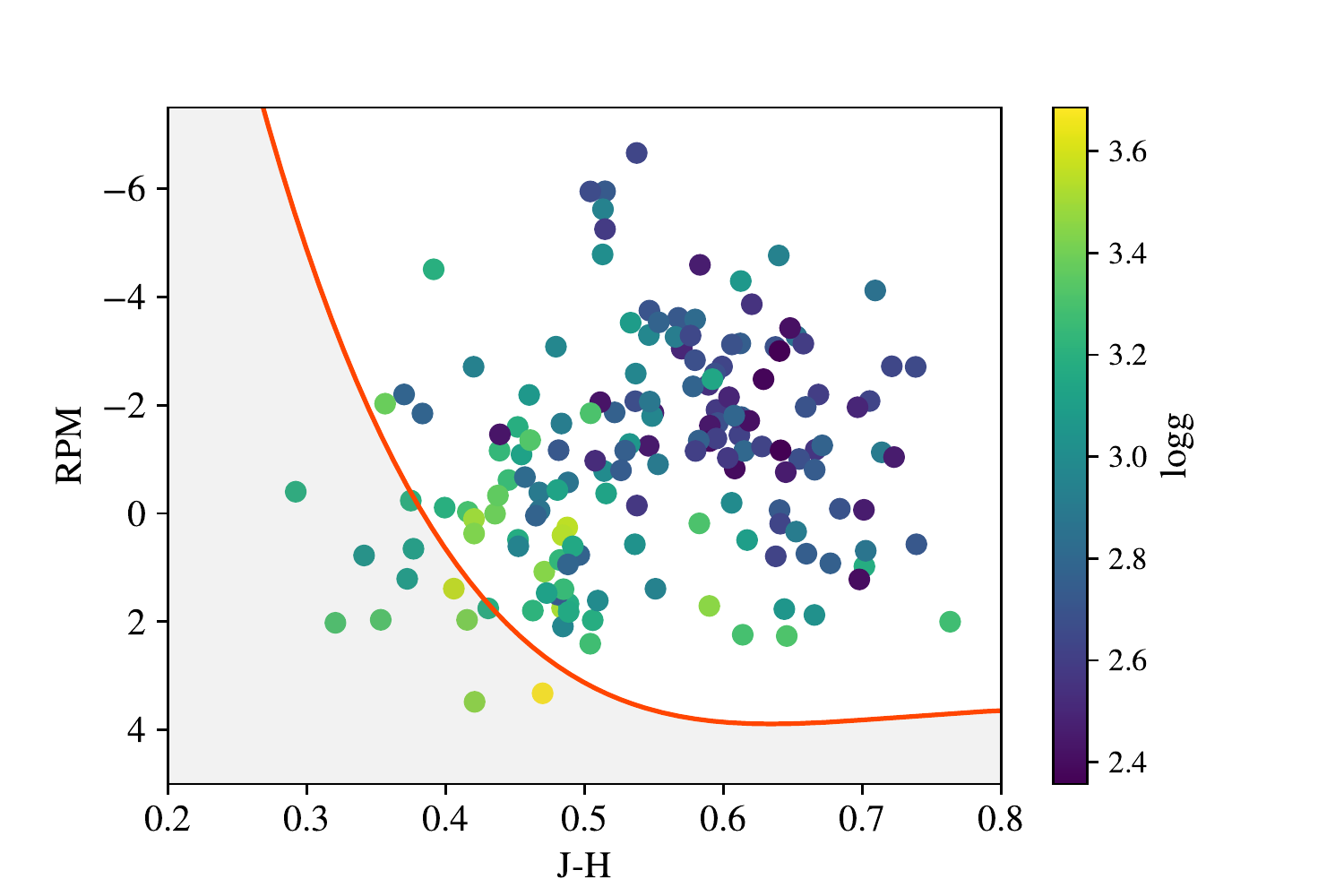}
      \caption{Giant star classification, using the prescription from \citet{CollierCameron2007}, described in Sect.~\ref{sec:initial_parameters}. The red line is the boundary between the dwarf and giant stars region. Points falling in the grey area are classified as dwarf stars.}
         \label{fig:classification}
   \end{figure}
   
   \begin{figure}
   \centering
   \includegraphics[width=\hsize]{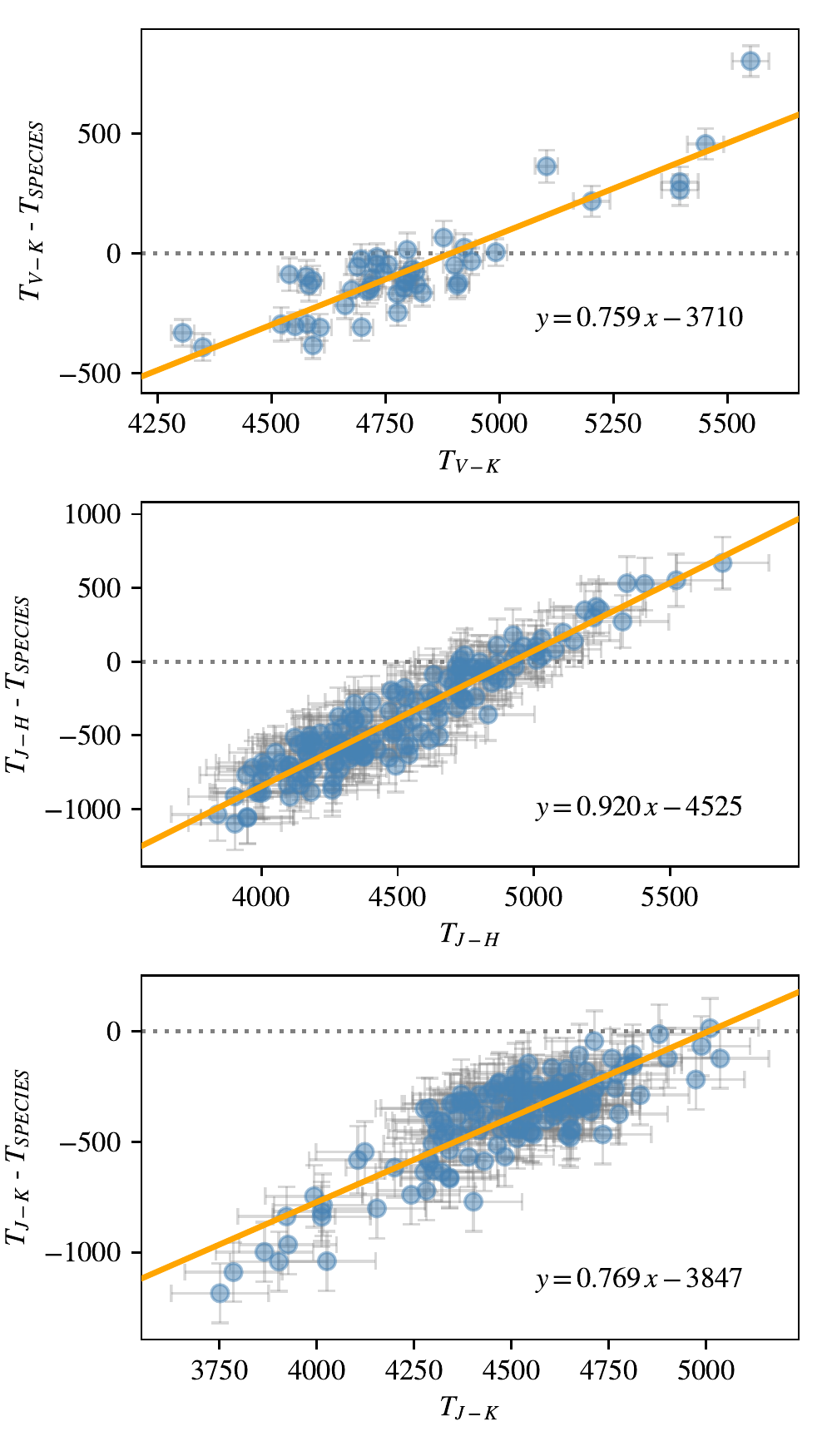}
      \caption{Comparison between temperature computed with the photometric relations from \citet{Alonso1999}, and SPECIES. Orange line represents the linear fit to the data, quoted in the text.}
         \label{fig:compare_T_photometry}
   \end{figure}

   \begin{figure}
   \centering
   \includegraphics[width=\hsize]{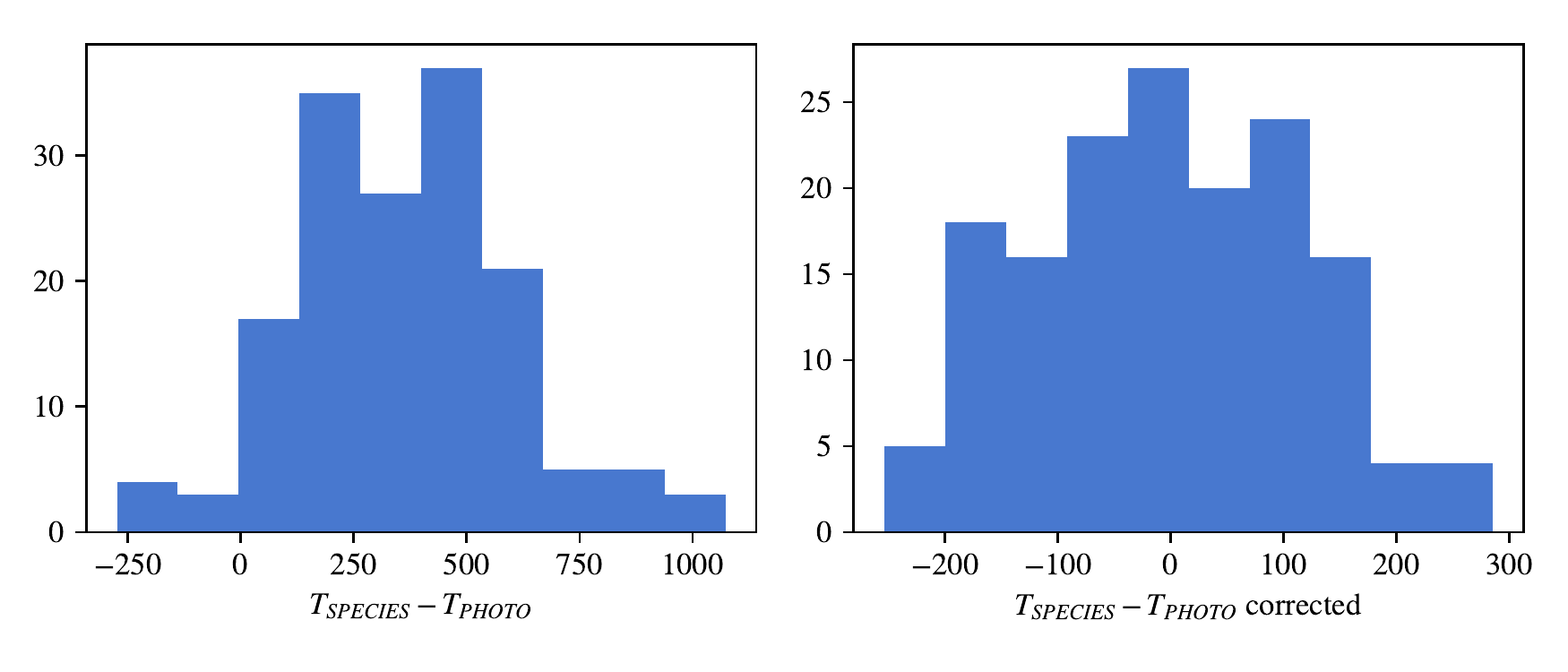}
      \caption{Histogram of the difference between the temperature from SPECIES, and from photometry, before (left panel) and after correcting (right panel) by the linear fits.}
         \label{fig:T_photo_corrected}
   \end{figure}

\begin{table}
    \caption{Linear fit parameters between the temperature obtained using the colour relations from \citet{Alonso1999} and SPECIES, with $T_{\text{eff,colour}} - T_{\text{eff,SPECIES}} = a\,T_{\text{eff,colour}}\, +\, b$.}
    \centering
    \begin{tabular}{l | l | l}
        \hline\hline
         Colour & $a$ & $b$  \\
         \hline
         V-K & 0.759 & 3710 \\
         J-H & 0.920 & 4525 \\
         J-K & 0.769 & 3847 \\
         \hline
    \end{tabular}
    \label{tab:coeffs_tphoto}
\end{table}

As was mentioned in Sect.~\ref{sec:initial_parameters}, we used the scheme from \citet{CollierCameron2007} to classify the stars processed with SPECIES as dwarf or giant stars. Figure~\ref{fig:classification} replicates Fig. 8 from \citet{CollierCameron2007}, but with the EXPRESS sample. We find that, following this scheme, 93\% of stars were correctly classified as giant. 

The initial temperature was derived from photometric information for each star, using the relations from \citet{Alonso1999}. Even though photometric relations are available for 12 colours within the paper, we only used the ones corresponding to $V-K$, $J-H$, and $J-K$. That is because we had very few stars with precise magnitude measurements in the remaining bands, and  therefore we could not test the reliability of the temperature relations for those colours. 
As is shown in \citet{Alonso1999}, the temperature relations also depend on the stellar metallicity. In the case of the initial parameters, we would have to use a photometric estimate of the metallicity. Most of the relations found in the literature use the Str\"{o}mgren photometric system, but only a small number of stars in our sample had those magnitudes available. Therefore, we decided to set the initial metallicity to zero for all the stars in the sample.

The individual temperatures using the relations from \citet{Alonso1999}, referred to as $T_{\text{eff,colour}}$, for [Fe/H] = 0, compared to the final values from SPECIES, are shown in Fig.~\ref{fig:compare_T_photometry}. We find that in all three cases, it is possible to relate both values through a linear fit of the form
\begin{equation}\label{eq:tphotocorr}
    T_{\text{eff,colour}} - T_{\text{eff,SPECIES}}\,=\,a\, T_{\text{eff,colour}}\,+\,b,
\end{equation}
with $a$ and $b$ listed in Table~\ref{tab:coeffs_tphoto}. The final temperature from photometry, referred as $T_{\text{eff,photo}}$, is taken as the average weight of the individual colour temperatures, with the weights given by the uncertainty in the temperature estimates, computed from the uncertainties in the magnitude measurements, and the intrinsic uncertainties from the colour temperature relations, given in \citet{Alonso1999}. 
In Fig.~\ref{fig:T_photo_corrected}, we show the histogram of the difference between the values from SPECIES with the final temperature average from photometry. In the case where no correction is applied to the individual colour relations, we find an offset of 358 K with respect to the SPECIES values, whereas if the correction is applied, that offset is reduced to 9 K.

We studied the effect of imposing [Fe/H] = 0 in the computation of $T_{\text{eff,photo}}$ by repeating the same analysis as before, but for two cases: setting [Fe/H] equal to the mean value found by SPECIES for the whole sample, -0.03 dex (Table~\ref{tab:mean_params_evol}), and using the metallicity computed by SPECIES for each individual star. We find that the parameters $a$ and $b$ from Eq.~\ref{eq:tphotocorr} do not change considerably, and that the mean difference between $T_{\text{eff,SPECIES}}-T_{\text{eff,photo}}$ changes at most by 5~K compared to the values from setting [Fe/H] = 0. We conclude that, for the stars in this sample, the assumption of [Fe/H]$_{\text{ini}} = 0$ does not affect the value for $T_{\text{ini}}$ from photometry. 

\subsection{Correlation between parameters}\label{sec:correlations}

   \begin{figure*}
   \centering
            \includegraphics[width=\textwidth]{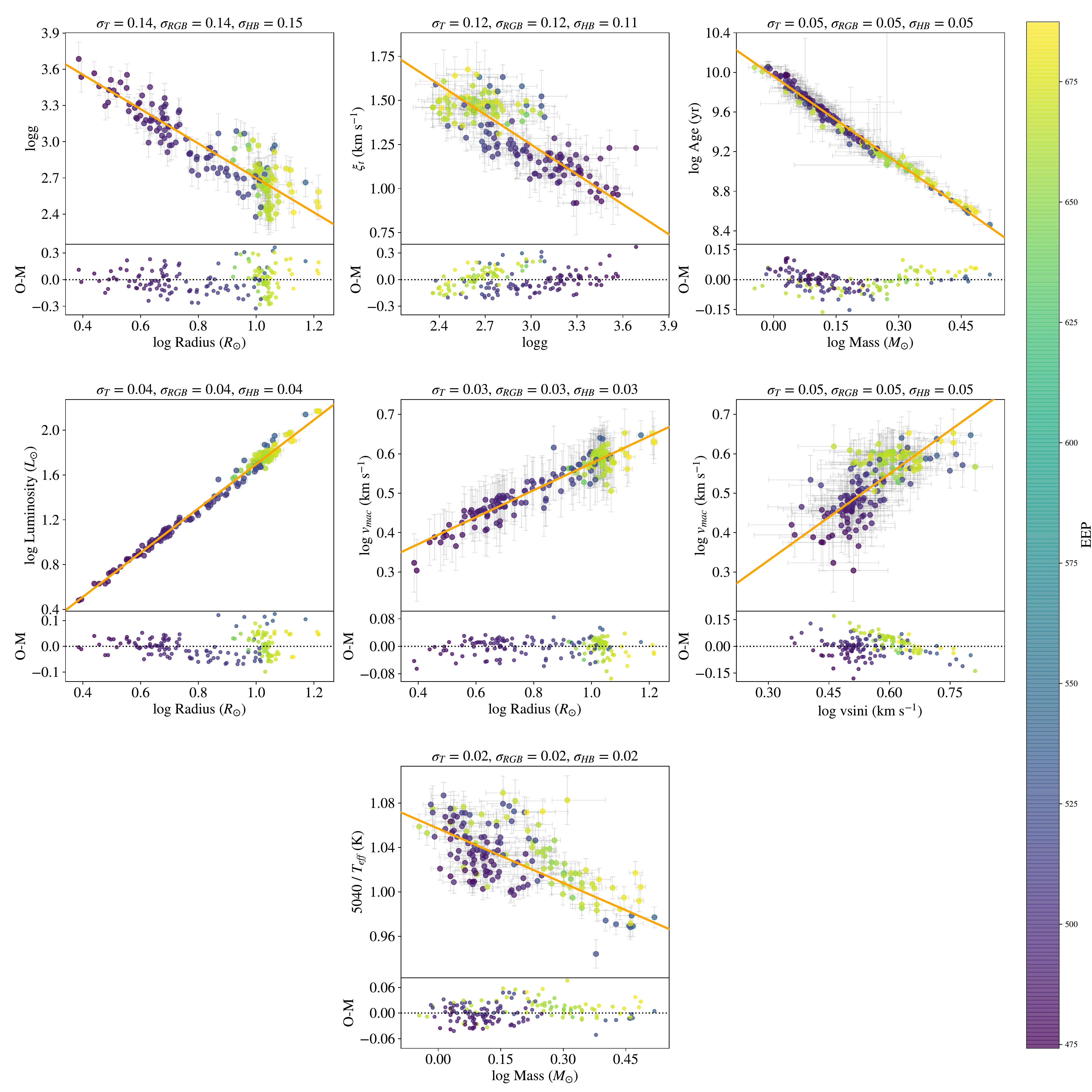}
      \caption{Correlations found between parameters derived in this work. The orange lines are the polynomial fits. Bottom panels are the residuals after subtracting the polynomial fits. The colour scale represents the median EEP for each target. Titles show the scatter in the residuals for the total sample ($\sigma_T$), stars in the RGB ($\sigma_{RGB}$) and in the HB ($\sigma_{HB}$).}
         \label{fig:correlations}
   \end{figure*}

   \begin{figure}
   \centering
   \includegraphics[width=\hsize]{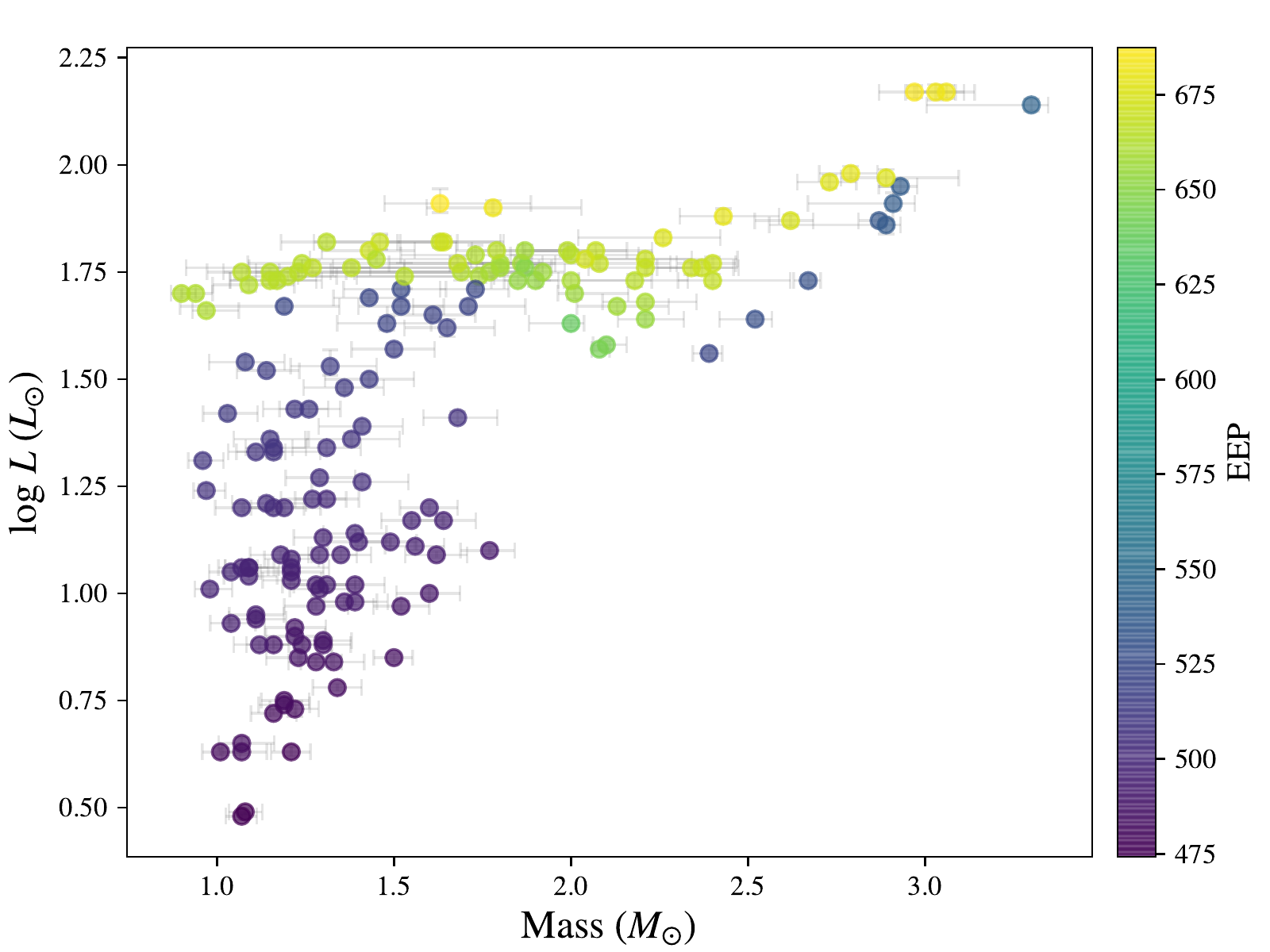}
      \caption{Mass-Luminosity diagram for the EXPRESS sample. The colour scale represents the median EEP for each target.}
         \label{fig:mass_logL_correlation}
   \end{figure}

We studied the degree of correlation between the parameters derived by SPECIES.
We estimated the Pearson correlation coefficients, $r$, between each parameter (Table~\ref{tab:correlations}), and consider the cases of clear correlation as the ones with $|r| > 0.5$.
Figure~\ref{fig:correlations} shows a selection of these strong correlations. Correlations for all the parameters are shown in Fig.~\ref{fig:correlations_all}, generated using the \texttt{corner}\footnote{\url{https://corner.readthedocs.io/en/latest/}} module. 
After fitting polynomial functions to the data, we derived Eqs.~\ref{eq:logg-radius}-\ref{eq:T-mass}.

\begin{align} 
\text{log\,g}\, &=  \phantom{-}(4.13 \pm 0.05) - (1.43 \pm 0.05)\log R\label{eq:logg-radius} \\ 
\xi_t\, &= \phantom{-}(2.95 \pm 0.09) - (0.57 \pm 0.03)\,\text{log\,g}\label{eq:vt-logg} \\
\log({\text{Age}})\, &= \phantom{-}(9.97 \pm 0.01) - (3.00 \pm 0.02)\log M\label{eq:age-mass}\\
\log\,L\, &= -(0.28 \pm 0.01) + (1.97 \pm 0.01)\log R\label{eq:luminosity-radius}\\
\log v_{mac}\, &= \phantom{-}(0.23 \pm 0.01) + (0.34 \pm 0.01)\log R\label{eq:vmac-radius}\\
\log v_{mac}\, &= \phantom{-}(0.11 \pm 0.03) + (0.74 \pm 0.05)\,\log v\sin i\label{eq:vmac-vsini}\\
\Theta\, &= \phantom{-}(1.06 \pm 0.01) - (0.16 \pm 0.01)\,\log M\label{eq:T-mass},
\end{align}

where $\Theta\,=\,5040/T_{\text{eff}}$.

The relation between the surface gravity and radius is expected by the physical definition of the surface gravity: $\text{g}\,=\, GM/R^2$, where $G$ is the gravitational constant, which can be rewritten as $\log\text{g}\,\sim\, a(M) - b\log R$, with $a(M)$ a value that depends on the mass. The mass range covered by the EXPRESS sample is for the most part very narrow (78\% of the stars have $0.85 M_{\odot} \leq M \leq 2 M_{\odot}$), which allows us to approximate $a(M)$ into a constant.

The relationship between microturbulence and surface gravity had been spotted before \citep{Monaco2005, Kirby2009, Jones2011, Adibekyan2015, Mucciarelli2020}, and takes the same shape as it does here ($\xi_t = a + b\,\text{log\,g}$). The coefficients we found ($a = 2.95$, $b = -0.57$) agree with the ones found in \citet{Jones2011}.

The age-mass relation is related to the fact that, once a star reaches the red giant phase, its age can be determined by the amount of time it spent in the main sequence \citep{Casagrande2016}. The time in the main sequence is directly related to the mass of the star, following $t_{MS} \propto  M^{-2.5}$. In our case, we find the power-law coefficient to be -3.0.

The luminosity is directly related to radius by its definition: $L \propto R^2\,T_{\text{eff}}^4$. As the range of temperatures mapped by the stars of the EXPRESS sample is narrow ($\sigma_{T_{\text{eff}}} = 128$ K), we can set the temperature dependence as a constant, which leads us to Eq.~\ref{eq:luminosity-radius}.

We derive Eqs.~\ref{eq:vmac-radius} and \ref{eq:vmac-vsini} that relate the macroturbulent velocity, the radius, and the rotational velocity. We see that the macroturbulent velocity saturates out at for the faster rotating stars, since the line profile is dominated by rotation in these cases, no matter the stellar spectral type.  The macroturbulence is driven by the stellar convective flows, and hence temperature, whereas the rotation velocity is tied to the evolution, and hence given the range of temperatures we are sensitive to, we can expect the rotation velocity to span a much wider range than the macroturbulent velocity.

Finally, Eq.~\ref{eq:T-mass} tells us that more massive stars also have larger surface temperatures, though the effect is more clear for core-helium burning stars, given that their position in the HB is determined by their mass (with more massive stars to the left of the HB).

From Table~\ref{tab:correlations} we can also see a large correlation coefficient between the stellar mass and luminosity. Both quantities are plotted in Fig.~\ref{fig:mass_logL_correlation}. 
For stars with $M < 2\,M_{\odot}$, the luminosity during the RGB will depend on the core mass, which explains the large spread seen in the diagram. During the HB, the luminosity of the star will be directly proportional to the helium core mass. After the helium-flash all stars will begin the HB with the same core mass, which explains why all HB stars have almost constant luminosity independent of mass. Stars with $M > 2\,M_{\odot}$ do not form a degenerate helium core, instead igniting helium quietly, which explains their different behaviour from lower-mass stars. The strong correlation seen in Table~\ref{tab:correlations} for mass and luminosity is driven mostly by the population with $M > 2\,M_{\odot}$, but these correspond to only 33 stars, therefore we can not confidently perform a regression fit.

The rest of the relations with large correlation coefficients (Table~\ref{tab:correlations}) can be inferred from the relations we have already presented (Eqs.~\ref{eq:logg-radius}-\ref{eq:T-mass}). For example, $v_{mac}$ is correlated with both stellar radius and luminosity, and luminosity and radius are correlated with each other as well, therefore it is possible to infer the relation between $v_{mac}$ and $L_{\star}$ from Eqs.~\ref{eq:luminosity-radius} and \ref{eq:vmac-radius}.

\subsection{Evolutionary stage}\label{sec:evolutionary_stage}

   \begin{figure}
   \centering
   \includegraphics[width=\hsize]{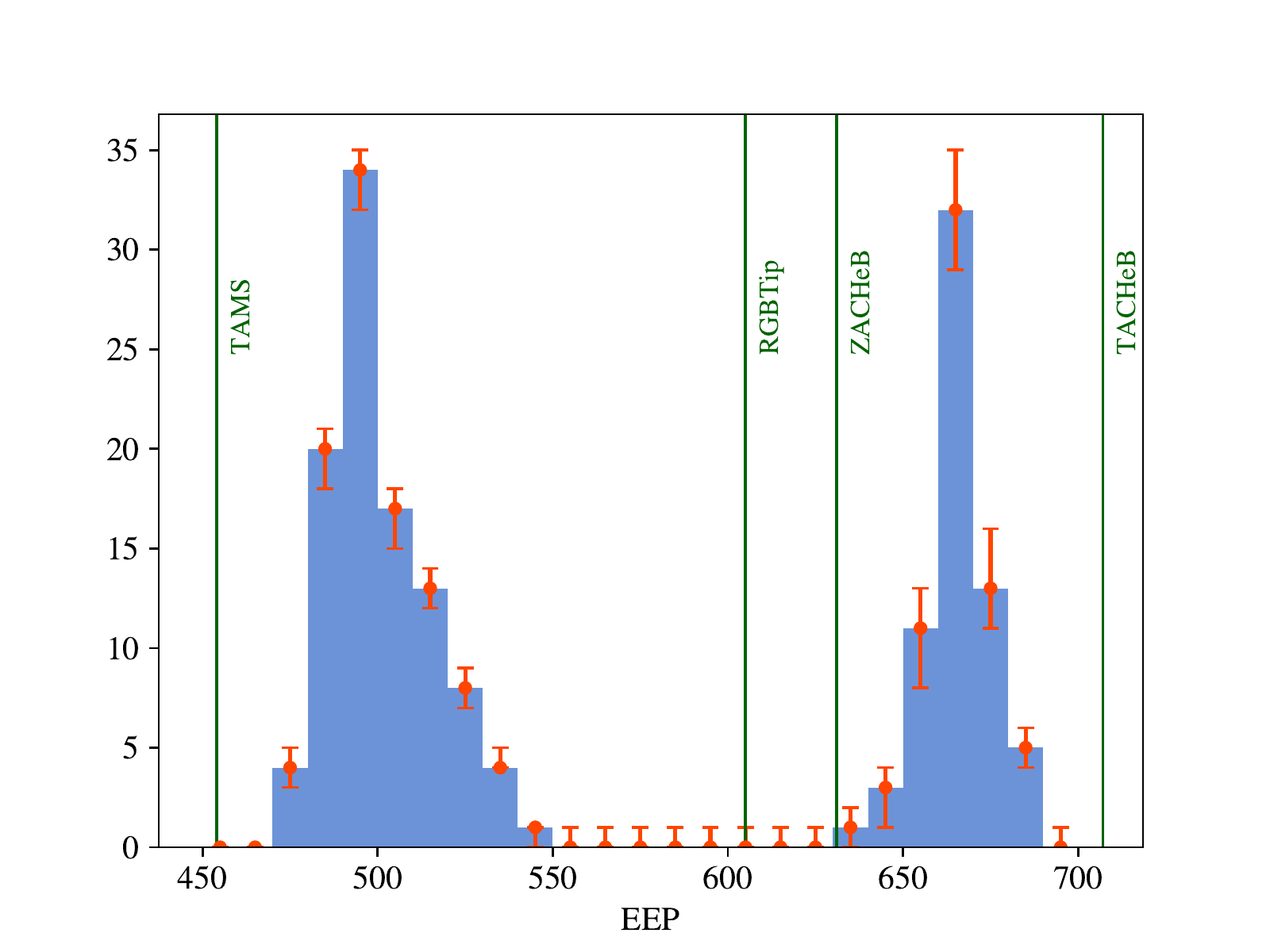}
      \caption{Distribution of stellar evolutionary stage for the EXPRESS sample. The errorbars in each bin reflect the uncertainty in the EEP values, given by the 16\% and 84\% percentiles of the EEP distribution for each star.}
         \label{fig:evolutionary_stage}
   \end{figure}
   
   \begin{figure*}
   \centering
            \includegraphics[width=\textwidth]{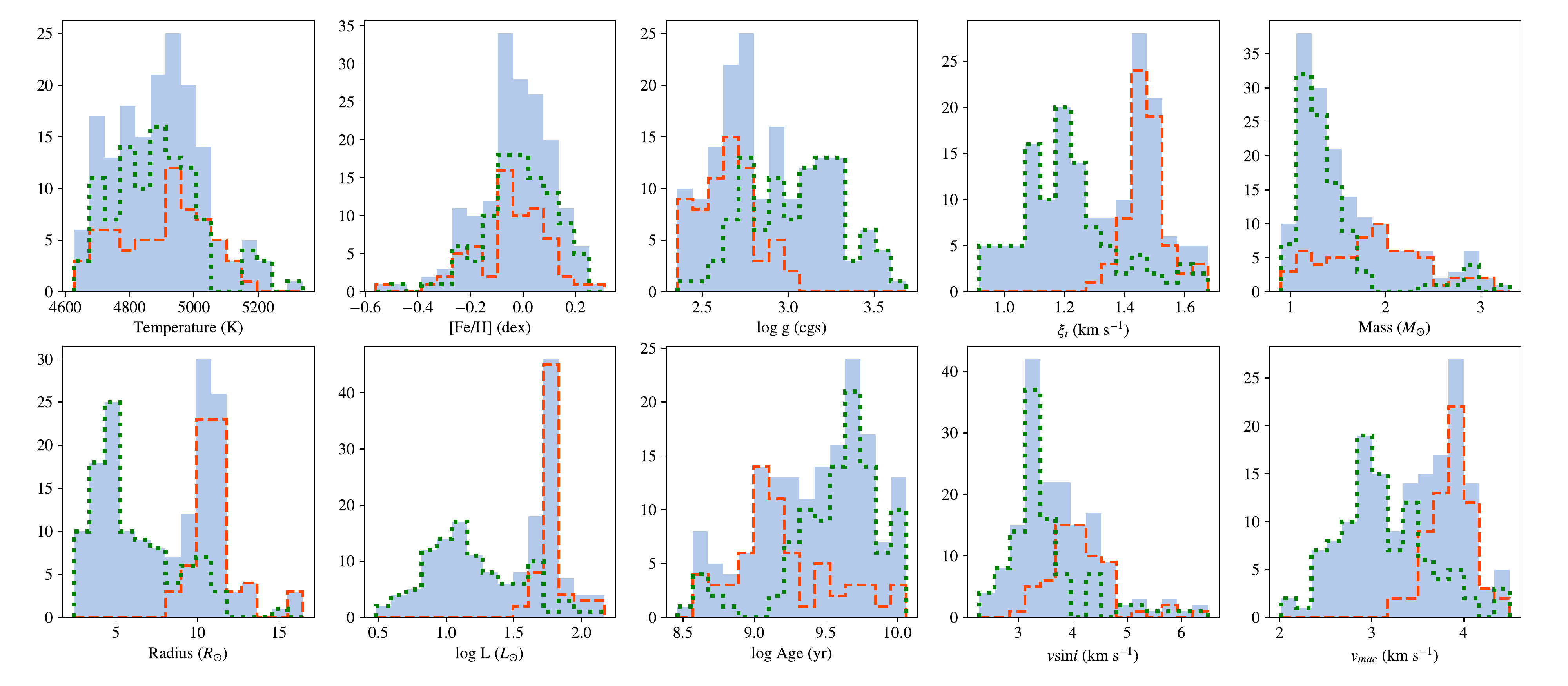}
      \caption{Distribution of some of the parameters derived by SPECIES, separated by their evolutionary stage. The red line corresponds to the HB distribution, and the green line to the RGB distribution. The blue histogram represents the total of both distributions.}
         \label{fig:histograms}
   \end{figure*}

\begin{table}
\caption{Mean and standard deviation for the parameters derived with SPECIES, and how they change with evolutionary stage.}
\label{tab:mean_params_evol}
\centering
\resizebox{.45\textwidth}{!}{%
\begin{tabular}{l | c  c  c}
\hline\hline
Parameter & Total & RGB & HB \\
 \hline
Temperature (K) & 4892 $\pm$ 138 & 4886 $\pm$ 138 & 4902 $\pm$ 139 \\
{[Fe/H]} (dex) & -0.03 $\pm$ 0.14 & -0.01 $\pm$ 0.13 & -0.05 $\pm$ 0.14 \\
log g (cgs) & 2.89 $\pm$ 0.31 & 3.05 $\pm$ 0.28 & 2.65 $\pm$ 0.17 \\
$\xi_t$ (km s$^{-1}$) & 1.31 $\pm$ 0.19 & 1.21 $\pm$ 0.16 & 1.47 $\pm$ 0.07 \\
Mass ($M_{\odot}$) & 1.59 $\pm$ 0.54 & 1.41 $\pm$ 0.46 & 1.87 $\pm$ 0.53 \\
Radius ($R_{\odot}$) & 7.99 $\pm$ 3.29 & 6.00 $\pm$ 2.49 & 11.07 $\pm$ 1.53 \\
log L ($L_{\odot}$) & 1.42 $\pm$ 0.40 & 1.19 $\pm$ 0.34 & 1.79 $\pm$ 0.11 \\
log Age (yr) & 9.43 $\pm$ 0.39 & 9.57 $\pm$ 0.34 & 9.20 $\pm$ 0.35 \\
$v\sin i$ (km s$^{-1}$) & 3.73 $\pm$ 0.76 & 3.48 $\pm$ 0.75 & 4.13 $\pm$ 0.60 \\
$v_{mac}$ (km s$^{-1}$) & 3.41 $\pm$ 0.55 & 3.11 $\pm$ 0.49 & 3.87 $\pm$ 0.24 \\
\hline
\end{tabular}}%
\end{table}

As mentioned in Sect.~\ref{sec:physical_parameters}, one of the outputs from \texttt{isochrones} is the EEP state for each star. We consider the final EEP value for a certain star to be the median of the EEP distribution, and follow the RGB and HB definition from Sect.~\ref{sec:physical_parameters}. Figure~\ref{fig:evolutionary_stage} shows the distribution of EEP values for our sample. We find that 61\% of stars (101) are in the RGB phase, while the other 39\% (65) are in the HB.

Figure~\ref{fig:histograms} shows the distribution of the stellar parameters as a function of evolutionary stage, with mean values separated into total, RGB, and HB distribution listed in Table~\ref{tab:mean_params_evol}. We find that there is no difference in the temperature and metallicity distributions between RGB and HB stars. We also find that the EXPRESS sample is slightly metal poor ($\overline{\text{[Fe/H]}}$ = -0.03 dex), which is similar to what has been seen in other radial velocity searches around evolved stars \citep{Dollinger2009, Mortier2013a}.

When looking at the size distribution, HB stars dominate the distribution for $R > 9\, R_{\odot}$, and in turn have, on average, larger radii than RGB stars. They are also distributed around $8.3 < R/R_{\odot} < 16.5$, whereas RGB stars can cover a larger range, with $2.4 < R/R_{\odot} < 15$. 
This radius difference between RGB and HB stars is reflected in the distribution of luminosity, surface gravity, micro and macroturbulence velocity, because of the correlations found in Sect.~\ref{sec:correlations}. We find that HB stars are more luminous, have on average smaller logg, and larger micro and macro turbulent velocity, than RGB stars. 

Low-mass stars ($M < 2 M_{\odot}$) are mostly found in the RGB phase (69\%), which is explained by the fact that the evolutionary timescale of this phase is longer for these stars than more massive stars, and therefore is more probable to find them at that stage. 
The near lack of RGB stars with masses $M > 2\,M_{\odot}$ is explained by the fact that intermediate mass stars do not go through the helium-flash (the ignition of helium burning in the stellar core from degenerate conditions) and therefore the lifetime of the RGB phase is much shorter than for lower-mass stars \citep{Girardi2013}.
HB stars also cover a large range of possible masses (from 0.9 $M_{\odot}$ to 3.06 $M_{\odot}$), but RGB stars are mostly clustered around 1.4 $M_{\odot}$. 
The dispersion of the residuals from the correlations adjusted in the previous section are large for HB stars in the case of the radius with logg. That could be explained by our approximation of the mass dependency on Eq.~\ref{eq:logg-radius} to a constant, which is better fitted for RGB stars with a narrow range of masses, than for HB stars.

Finally, we find that HB stars have on average larger rotational velocities than RGB stars.

\section{Summary and conclusions}\label{sec:summary}

In this work we present atmospheric and physical parameters of the EXPRESS program sample, derived using the SPECIES code. 
In this work we have introduced some improvements to SPECIES, including an updated version of the line list, a new routine to measure EWs of the iron lines and the addition of the EEP into the interpolation of evolutionary models. Similarly, here we use higher quality stellar spectra, more accurate parallaxes from the Gaia DR2 and newer extinction maps, compared to the original catalogue. Based on the posterior probabilities, we find that 101 stars are most probably in the RGB, and 65 in the HB. The separation of both stages is important not only for breaking the degeneracy in the H-R diagram and the correct estimation of the stellar mass, but also for the study of the effect of stellar evolution in planetary systems.
\newline \indent
We compare our results with asteroseismology studies. Overall we find an agreement at the 2\% level in the stellar radii and at the 6\% level in the stellar mass. By comparing our stellar radii with interferometric studies we also find an agreement better than 2\%. 
These results validate our method and show the robustness of SPECIES to accurately derived stellar parameters. \newline \indent
Similarly, we compared our temperature, surface gravity, metallicity, mass, radius, and stellar luminosity with estimates from other spectroscopic catalogues. We find good agreement for all the parameters, with mean differences within 2\%. The only exception is the mass, for which we find an overall difference of -0.15 $M_{\odot}$, corresponding to $\sim$ 9\% difference. The largest differences are with respect to  J11. This could be due to the differences in the line list, the stellar evolution models and interpolation procedure, the quality of the spectra used, or to the complexity of estimating stellar masses for giant stars using stellar evolution tracks, as these are degenerate in the parameter space giant stars occupy. \newline \indent 
In addition, we studied the correlations between the parameters and find six relations, most of them driven by physical processes (logg-radius, age-mass, and luminosity-radius) and others already detected in previous works (microtrubulence-logg). We also detect relations between the macroturbulence velocity and the radius, and between the macroturbulence and rotational velocity. \newline \indent
Finally we compared our spectroscopic and photometric logg's, to investigate for any potential systematic difference, as has been widely reported in the literature. We find an excellent agreement between these two quantities, with a mean difference of 0.001 cgs. Again this result shows the internal consistency and robustness of the method.

\begin{acknowledgements}

MGS acknowledges support from STFC through the Consolidated Grant ST/M001202/1.
JSJ acknowledges support by FONDECYT grant 1201371 and partial support from CONICYT project Basal AFB-170002.
This research has made use of the VizieR catalogue access tool, CDS,
 Strasbourg, France (DOI : 10.26093/cds/vizier). The original description 
 of the VizieR service was published in 2000, A\&AS 143, 23.
 
\end{acknowledgements}

\bibliographystyle{aa}
\bibliography{speciesii}

\begin{appendix}

\section{Equivalent Width computation}\label{sec:EW}

The equivalent widths (EWs) were computed using the \texttt{EWComputation} module inside SPECIES\footnote{Also available on its own at \url{https://github.com/msotov/EWComputation}}. For each line at wavelength $l$ in a given linelist, this module: 1) selects a region within 3 \angstrom\, from the line and normalises the spectrum; 2) detects the absorption lines within the region; 3) fits a Gaussian-like profile to the line; and 4) computes the equivalent width of the line based on the fit parameters, and its corresponding uncertainty. Steps 1) and 2) were written following the prescription from \citet{Sousa2007}. 

\subsection{Continuum normalization}
The continuum normalization is done by fitting 2nd degree polynomials to the data, and rejecting the points that lie further than {\it rejt} from the polynomial. The value {\it rejt} was first introduced in \citet{Sousa2007}, and is defined in \texttt{SPECIES} as {\it rejt} = 1-1/SNR, where SNR is the signal-to-noise ratio of the line region. The SNR is typically taken from the headers of the spectra, and it can be a global value, representing the whole spectrum, or have different values depending on the wavelength of the data. 

\subsection{Line detection}
The detection of spectral lines is done by identifying the regions in the spectrum where the derivative is zero, and where the 2nd degree derivative is positive, indicating that the point corresponds to a local minimum. Before doing the derivations, the spectrum is smoothed by taking the convolution with a Box Kernel of size 4, and only the points with flux < -0.02 are considered as spectral lines. This is done to deal with very noisy data, where fluctuations in the continuum can be mistaken as line profiles. This will result in a set of $N$ lines centred at wavelengths $\mu_i$, where $i=1,...,N$. We additionally define the index $I$ which refers to the line closest to the target point, as defined by the criterion $\mu_I = \mbox{MIN}\{|\mu_i-l|\}$.

\subsection{Line profile fit}

Once the $N$ absorption lines are identified, we assume that each line follows a Gaussian-shaped profile, centered at wavelengths $\mu_i$. The spectrum $S$ at wavelength $w$ can then be written as:
\begin{equation}
S(w)\,=\, \sum_{i=1}^N\, A_i \exp{[-(w-\mu_i)^2/2\sigma_i^2]}.
\end{equation}

We use the Scipy implementation of the Ortogonal Distance Regression \citep[ODR, ][]{odr} method\footnote{\url{https://docs.scipy.org/doc/scipy/reference/odr.html}} to minimize the above expression and estimate the values $A_i$, $\mu_i$, and $\sigma_i$ that best represent each line, with its corresponding uncertainties $e_{A_i}$, $e_{\mu_i}$, and $e_{\sigma_i}$. The fit for the central line $I$ is accepted if $w_I-l \leq 0.075$ \angstrom, $A_I \leq 0$ (absorption feature), $\sigma_I < 0.15$ \angstrom, and the uncertainties for all the parameters is less than 0.12 dex. Otherwise, the line is discarded from the computation. These restrictions help us with the detection and fitting of blended lines.  

\begin{figure}
   \centering
   \includegraphics[width=\hsize]{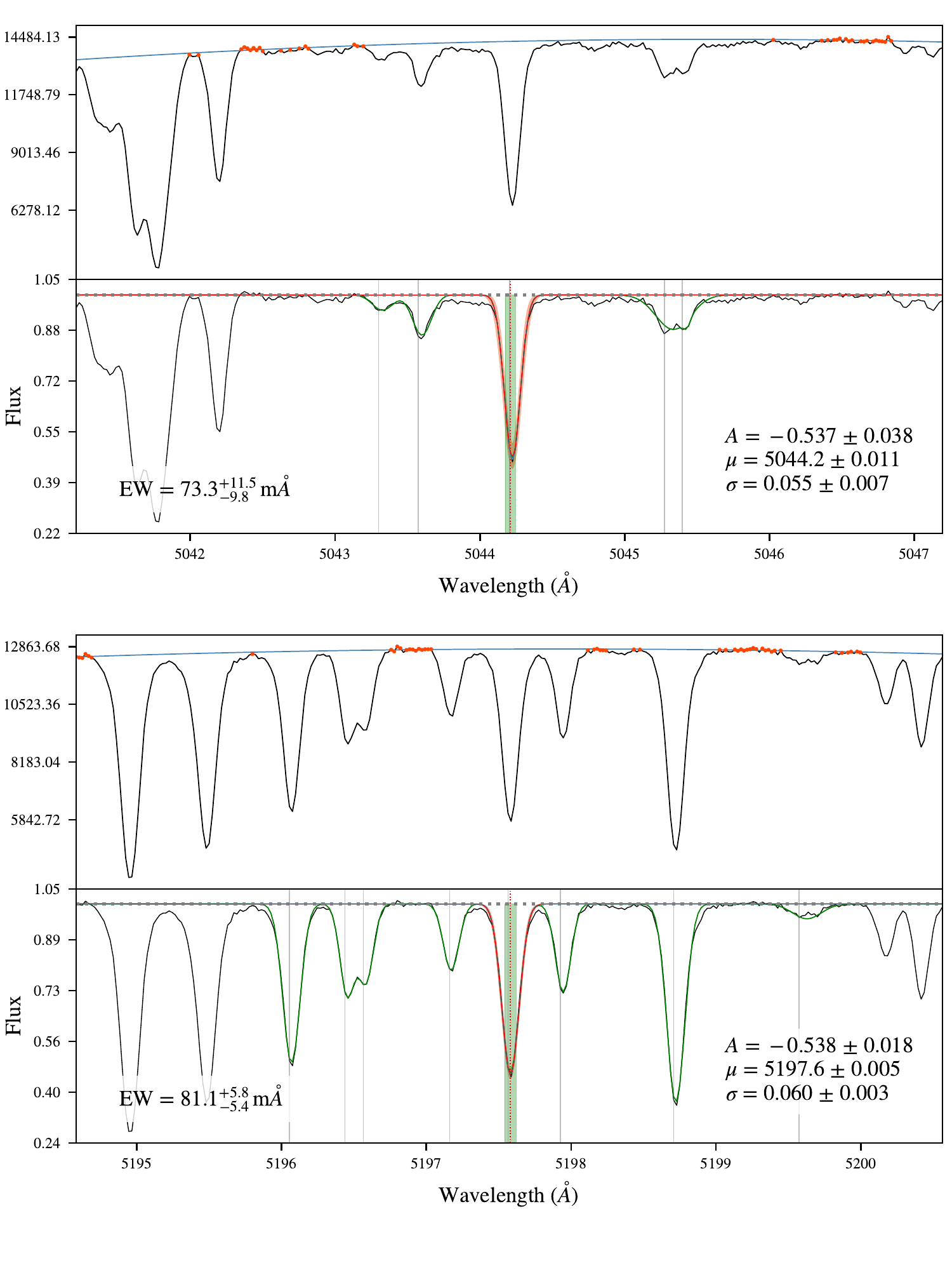}
      \caption{Fit performed to the \FeI 5044.21 \angstrom ~and \FeII 5197.58 \angstrom  ~lines from one of the Sun spectra obtained by observing Ceres with HIRES.
      The top panels in both plots show the continuum fitting procedure, where the red dots are the final points selected for the fitting of the 2nd degree polynomial (blue line).
      The bottom panels show the line fitting procedure.
      The gray lines are the absorption lines detected in the spectral range, and the green line the global fit to the data. The red region corresponds to the fit of the line $l$ with its uncertainty, derived using the Gaussian parameters plus their uncertainties (quoted in the text). The green block represents the equivalent width.}
         \label{fig:ceres01_lines}
   \end{figure}

Line profiles fitted using our method are shown in Fig.~\ref{fig:ceres01_lines}.

\subsection{Equivalent width estimation}

For each line $l$, the equivalent width is computed by integrating the line profile with parameters $A_I$, $w_I$, and $\sigma_I$, over the spectral region surrounding the line. The uncertainty in this value is estimated by changing the Gaussian parameters of the fit within 1$\sigma$ of their uncertainty, and then selecting the 16\% and 84\% percentiles of the distribution.

   \begin{figure*}
   \centering
            \includegraphics[width=\textwidth]{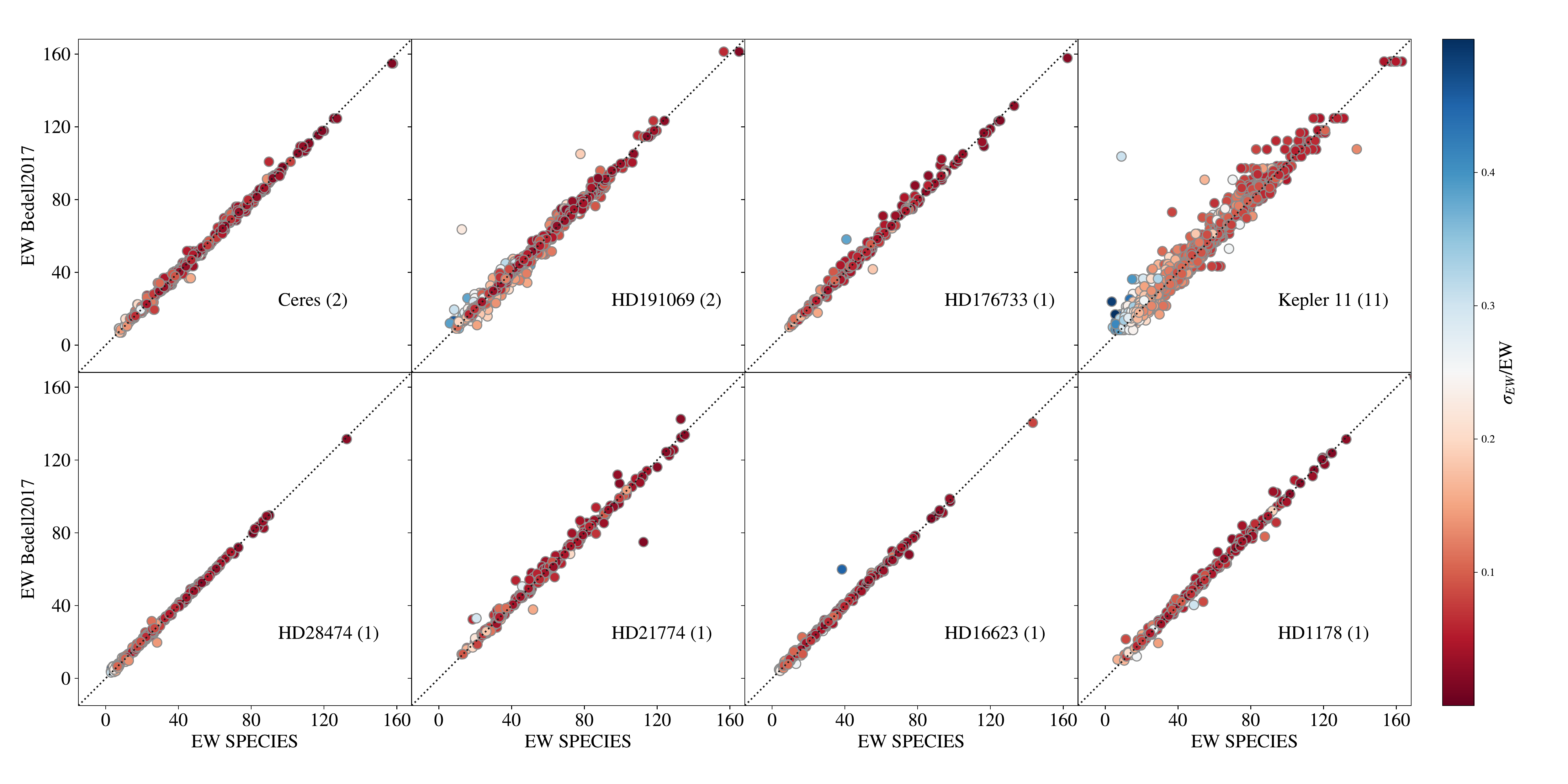}
      \caption{Comparison between the equivalent widths computed using SPECIES, against the values from \citet{Bedell2017}. The number next to the names corresponds to the number of spectra available for each star. The color represents the uncertainty in the EW obtained with our method.}
              
         \label{fig:compare_EW}
   \end{figure*}

\begin{table}
\label{tab:compare_EW}
\caption{Mean ($\mu$), error in the mean ($\sigma_{\mu}$), and standard deviation ($\sigma$) of the difference between the EWs computed by SPECIES, and from \citet{Bedell2017}.}
\centering
\begin{tabular}{lccc}
\hline\hline
Star name & $\mu$ & $\sigma_{\mu}$ & $\sigma$ \\
& (m\AA) & (m\AA) & (m\AA) \\
\hline
Ceres & -0.19 & 1.16 & 1.69 \\
HD 191069 & -0.14 & 2.78 & 4.11 \\
HD 176733 & -0.07 & 2.61 & 2.54 \\
Kepler 11 & 0.09 & 2.73 & 4.29 \\
HD 28474 & 0.08 & 0.73 & 1.09 \\
HD 21774 & -0.27 & 2.44 & 3.69 \\
HD 16623 & 0.26 & 1.83 & 1.80 \\
HD 1178 & 0.02 & 1.66 & 2.41 \\
\hline
\end{tabular}
\end{table}

In order to assess the reliability of our method, we used it to estimate the equivalent width of the Sun, plus other 7 stars considered to be solar twins, and compared them to the values obtained by \citet{Bedell2017}, where the continuum estimation and line profile fitting were performed for each line by hand using the \textit{splot} module of IRAF\footnote{IRAF is distributed by the National Optical Astronomy Observatories,
    which are operated by the Association of Universities for Research
    in Astronomy, Inc., under cooperative agreement with the National
    Science Foundation.}.
The results of out comparison, for the 8 stars, is shown in Fig.~\ref{fig:compare_EW}, and Table~\ref{tab:compare_EW}. The spectra were taken with the HIRES spectrograph \citep{Vogt1994}, and we used the same line list as in \citet{Bedell2017}.

\section{Supplementary figures and tables}

\begin{table*}
\caption{Line list used in the atmospheric parameters derivation.}
    \label{tab:linelist}
    \centering
    \begin{tabular}{cccc | cccc}
    \hline\hline
$\lambda$ (\AA) & $\chi_l$ & $\log gf$ & Element & $\lambda$ (\AA) & $\chi_l$ & $\log gf$ & Element \\
\hline
5494.47 & 4.07 & -1.96 & \FeI & 6151.62 & 2.18 & -3.26 & \FeI \\
5522.45 & 4.21 & -1.47 & \FeI & 6157.73 & 4.07 & -1.26 & \FeI \\
5539.29 & 3.64 & -2.59 & \FeI & 6165.36 & 4.14 & -1.48 & \FeI \\
5560.22 & 4.43 & -1.1 & \FeI & 6173.34 & 2.22 & -2.84 & \FeI \\
5608.98 & 4.21 & -2.31 & \FeI & 6188.0 & 3.94 & -1.6 & \FeI \\
5611.36 & 3.63 & -2.93 & \FeI & 6200.32 & 2.61 & -2.39 & \FeI \\
5618.64 & 4.21 & -1.34 & \FeI & 6213.44 & 2.22 & -2.54 & \FeI \\
5619.61 & 4.39 & -1.49 & \FeI & 6219.29 & 2.2 & -2.39 & \FeI \\
5635.83 & 4.26 & -1.59 & \FeI & 6226.74 & 3.88 & -2.08 & \FeI \\
5650.0 & 5.1 & -0.8 & \FeI & 6232.65 & 3.65 & -1.21 & \FeI \\
5652.33 & 4.26 & -1.77 & \FeI & 6240.65 & 2.22 & -3.23 & \FeI \\
5701.56 & 2.56 & -2.16 & \FeI & 6246.33 & 3.6 & -0.73 & \FeI \\
5717.84 & 4.28 & -0.98 & \FeI & 6252.57 & 2.4 & -1.64 & \FeI \\
5731.77 & 4.26 & -1.1 & \FeI & 6265.14 & 2.18 & -2.51 & \FeI \\
5738.24 & 4.22 & -2.24 & \FeI & 6270.23 & 2.86 & -2.55 & \FeI \\
5741.86 & 4.26 & -1.69 & \FeI & 6280.62 & 0.86 & -4.34 & \FeI \\
5752.04 & 4.55 & -0.92 & \FeI & 6297.8 & 2.22 & -2.7 & \FeI \\
5760.36 & 3.64 & -2.46 & \FeI & 6301.51 & 3.65 & -0.72 & \FeI \\
5775.09 & 4.22 & -1.11 & \FeI & 6311.5 & 2.83 & -3.16 & \FeI \\
5778.46 & 2.59 & -3.44 & \FeI & 6315.81 & 4.07 & -1.67 & \FeI \\
5793.92 & 4.22 & -1.62 & \FeI & 6322.69 & 2.59 & -2.38 & \FeI \\
5806.73 & 4.61 & -0.93 & \FeI & 6330.85 & 4.73 & -1.22 & \FeI \\
5814.82 & 4.28 & -1.81 & \FeI & 6335.34 & 2.2 & -2.28 & \FeI \\
5835.11 & 4.26 & -2.18 & \FeI & 6380.75 & 4.19 & -1.34 & \FeI \\
5837.7 & 4.29 & -2.3 & \FeI & 6481.88 & 2.28 & -2.94 & \FeI \\
5859.6 & 4.55 & -0.63 & \FeI & 6498.95 & 0.96 & -4.66 & \FeI \\
5862.37 & 4.55 & -0.42 & \FeI & 6518.37 & 2.83 & -2.56 & \FeI \\
5902.48 & 4.59 & -1.86 & \FeI & 6533.94 & 4.56 & -1.28 & \FeI \\
5905.68 & 4.65 & -0.78 & \FeI & 6574.25 & 0.99 & -4.96 & \FeI \\
5927.8 & 4.65 & -1.07 & \FeI & 6581.22 & 1.48 & -4.68 & \FeI \\
5929.68 & 4.55 & -1.16 & \FeI & 6593.88 & 2.43 & -2.3 & \FeI \\
5930.19 & 4.65 & -0.34 & \FeI & 6608.04 & 2.28 & -3.96 & \FeI \\
5934.67 & 3.93 & -1.08 & \FeI & 6609.12 & 2.56 & -2.65 & \FeI \\
5956.71 & 0.86 & -4.56 & \FeI & 6627.56 & 4.55 & -1.5 & \FeI \\
5976.79 & 3.94 & -1.3 & \FeI & 6633.76 & 4.56 & -0.81 & \FeI \\
5984.83 & 4.73 & -0.29 & \FeI & 6667.72 & 4.58 & -2.1 & \FeI \\
6003.02 & 3.88 & -1.02 & \FeI & 6703.58 & 2.76 & -3.0 & \FeI \\
6007.97 & 4.65 & -0.76 & \FeI & 6713.75 & 4.79 & -1.41 & \FeI \\
6008.57 & 3.88 & -0.92 & \FeI & 6725.36 & 4.1 & -2.21 & \FeI \\
6027.06 & 4.07 & -1.2 & \FeI & 6726.67 & 4.61 & -1.05 & \FeI \\
6056.01 & 4.73 & -0.46 & \FeI & 6733.15 & 4.64 & -1.44 & \FeI \\
6065.49 & 2.61 & -1.49 & \FeI & 6739.52 & 1.56 & -4.85 & \FeI \\
6078.5 & 4.79 & -0.38 & \FeI & 6745.97 & 4.07 & -2.71 & \FeI \\
6079.02 & 4.65 & -0.97 & \FeI & 6750.16 & 2.42 & -2.58 & \FeI \\
6082.72 & 2.22 & -3.53 & \FeI & 5197.58 & 3.23 & -2.23 & \FeII \\
6089.57 & 5.02 & -0.87 & \FeI & 5234.63 & 3.22 & -2.22 & \FeII \\
6093.65 & 4.61 & -1.32 & \FeI & 5991.38 & 3.15 & -3.55 & \FeII \\
6094.38 & 4.65 & -1.56 & \FeI & 6369.46 & 2.89 & -4.21 & \FeII \\
6096.67 & 3.98 & -1.76 & \FeI & 6416.93 & 3.89 & -2.7 & \FeII \\
6098.25 & 4.56 & -1.81 & \FeI & 6456.39 & 3.9 & -2.1 & \FeII \\
6137.0 & 2.2 & -2.91 & \FeI & 6516.08 & 2.89 & -3.38 & \FeII \\
\hline
\end{tabular}
\end{table*}

\begin{table*}
\caption{Parameters derived with SPECIES for the stars included in interferometric and asteroseismology works. Some stars in the list do not belong to the EXPRESS program, but were included just to test the method.}
\label{tab:test_SPECIES}
\centering
\begin{tabular}{l l c c c c c}
\hline\hline
HIP & HD & logg & Radius & Mass & Age & References\\
& & (cgs) & ($R_{\odot}$) & ($M_{\odot}$) & (Gyr) &\\
\hline
343 & 225197 & 2.44 $\pm$ 0.12 & 11.21$^{+0.22}_{-0.13}$ & 1.38$^{+0.15}_{-0.14}$ & 3.26$^{+1.21}_{-0.86}$ & 7 \\
655 & 344 & 2.36 $\pm$ 0.11 & 11.25$^{+0.41}_{-0.17}$ & 1.27$^{+0.16}_{-0.18}$ & 4.24$^{+2.49}_{-1.28}$ & 7 \\
3137 & 3750 & 2.38 $\pm$ 0.13 & 9.97$^{+0.09}_{-0.08}$ & 1.61$^{+0.12}_{-0.13}$ & 1.99$^{+0.55}_{-0.38}$ & 1 \\
4293 & 5457 & 2.69 $\pm$ 0.09 & 9.07$^{+0.24}_{-0.22}$ & 1.50$^{+0.12}_{-0.12}$ & 2.47$^{+0.73}_{-0.51}$ & 7 \\
4587 & 5722 & 2.69 $\pm$ 0.08 & 10.58$^{+0.31}_{-0.31}$ & 2.07$^{+0.16}_{-0.21}$ & 1.11$^{+0.17}_{-0.23}$ & 1 \\
8102 & 10700 & 4.33 $\pm$ 0.06 & 0.83$^{+0.01}_{-0.01}$ & 0.77$^{+0.01}_{-0.01}$ & 13.04$^{+0.34}_{-0.73}$ & 3, 5 \\
8928 & 11977 & 2.72 $\pm$ 0.08 & 10.64$^{+0.23}_{-0.20}$ & 2.29$^{+0.15}_{-0.12}$ & 0.82$^{+0.15}_{-0.17}$ & 1 \\
9440 & 12438 & 2.65 $\pm$ 0.08 & 9.82$^{+0.29}_{-0.23}$ & 1.86$^{+0.08}_{-0.03}$ & 1.09$^{+0.05}_{-0.04}$ & 1 \\
10234 & 13468 & 2.64 $\pm$ 0.08 & 10.57$^{+0.32}_{-0.29}$ & 2.00$^{+0.19}_{-0.17}$ & 1.19$^{+0.15}_{-0.23}$ & 1 \\
11791 & 15779 & 2.75 $\pm$ 0.10 & 9.90$^{+0.27}_{-0.24}$ & 2.18$^{+0.15}_{-0.16}$ & 1.03$^{+0.20}_{-0.23}$ & 1 \\
13147 & 17652 & 2.56 $\pm$ 0.08 & 10.15$^{+0.11}_{-0.10}$ & 1.87$^{+0.04}_{-0.96}$ & 1.16$^{+8.17}_{-0.02}$ & 1 \\
16537 & 22049 & 4.40 $\pm$ 0.06 & 0.72$^{+0.00}_{-0.00}$ & 0.80$^{+0.01}_{-0.01}$ & 0.59$^{+0.92}_{-0.38}$ & 3 \\
19849 & 26965 & 4.33 $\pm$ 0.06 & 0.77$^{+0.03}_{-0.02}$ & 0.75$^{+0.02}_{-0.02}$ & 12.18$^{+0.99}_{-2.15}$ & 3 \\
69673 & 124897 & 1.67 $\pm$ 0.10 & 24.76$^{+1.28}_{-1.13}$ & 0.91$^{+0.09}_{-0.06}$ & 9.87$^{+2.42}_{-2.67}$ & 2 \\
79672 & 146233 & 4.36 $\pm$ 0.05 & 1.04$^{+0.02}_{-0.02}$ & 1.03$^{+0.03}_{-0.03}$ & 3.92$^{+1.58}_{-1.56}$ & 3 \\
89962 & 168723 & 3.01 $\pm$ 0.08 & 5.54$^{+0.07}_{-0.06}$ & 1.57$^{+0.08}_{-0.08}$ & 1.89$^{+0.29}_{-0.23}$ & 2 \\
92968 & 175679 & 2.88 $\pm$ 0.10 & 12.63$^{+0.49}_{-0.48}$ & 3.04$^{+0.07}_{-0.26}$ & 0.35$^{+0.10}_{-0.02}$ & 2 \\
95124 & 181342 & 3.13 $\pm$ 0.10 & 5.31$^{+0.06}_{-0.06}$ & 1.64$^{+0.09}_{-0.09}$ & 2.00$^{+0.37}_{-0.27}$ & 8 \\
95222 & 181907 & 2.43 $\pm$ 0.10 & 11.20$^{+0.22}_{-0.13}$ & 1.45$^{+0.17}_{-0.14}$ & 2.68$^{+0.93}_{-0.74}$ & 2 \\
98036 & 188512 & 3.61 $\pm$ 0.07 & 3.00$^{+0.02}_{-0.02}$ & 1.37$^{+0.03}_{-0.03}$ & 3.02$^{+0.14}_{-0.11}$ & 2, 5 \\
102014 & 196737 & 2.71 $\pm$ 0.11 & 9.89$^{+0.32}_{-0.29}$ & 1.48$^{+0.12}_{-0.14}$ & 2.61$^{+0.96}_{-0.58}$ & 7 \\
103836 & 200073 & 2.75 $\pm$ 0.09 & 7.84$^{+0.21}_{-0.20}$ & 1.03$^{+0.09}_{-0.07}$ & 8.79$^{+2.32}_{-2.04}$ & 7 \\
105854 & 203949 & 2.46 $\pm$ 0.16 & 10.75$^{+0.22}_{-0.16}$ & 0.97$^{+0.09}_{-0.04}$ & 11.50$^{+1.40}_{-2.58}$ & 4, 7 \\
110813 & 212771 & 3.38 $\pm$ 0.07 & 4.33$^{+0.04}_{-0.04}$ & 1.57$^{+0.06}_{-0.07}$ & 1.95$^{+0.25}_{-0.17}$ & 4, 8 \\
114775 & 219263 & 2.70 $\pm$ 0.12 & 8.62$^{+0.27}_{-0.25}$ & 1.43$^{+0.13}_{-0.12}$ & 3.26$^{+1.16}_{-0.78}$ & 7 \\
116630 & 222076 & 3.17 $\pm$ 0.12 & 4.28$^{+0.05}_{-0.05}$ & 1.28$^{+0.09}_{-0.09}$ & 4.54$^{+1.20}_{-0.92}$ & 6 \\
\hline
\end{tabular}
\tablebib{
(1)~\citet{Gallenne2018}; (2) \citet{Morel2014}; (3) \citet{Tsantaki2013}; (4) \citet{Campante2019}; 
(5) \citet{Rains2020}; (6) \citet{Jiang2020}; (7) \citet{Aguirre2020}; (8) \citet{North2017}.
}
\end{table*}

\begin{table*}
\caption{Literature data for stars used in the comparison with interferometry and asteroseismology. Only the quantities plotted in Figs.~\ref{fig:interferometry} and~\ref{fig:asteroseismology} are listed.}\label{tab:test_literature}
\centering
\begin{tabular}{lccccc}
\hline\hline
HIP  & log g & Radius & Mass & Age & Reference \\
& (cgs) & ($R_{\odot}$) & ($M_{\odot}$) & (Gyr) & \\
\hline
343 & 2.54 $\pm$ 0.07\tablefootmark{a} & 10.69 $\pm$ 0.52 & 1.44 $\pm$ 0.18 & 3.50 $\pm$ 1.05 & 1 \\
655 & 2.41 $\pm$ 0.09\tablefootmark{a} & 11.27 $\pm$ 0.77 & 1.19 $\pm$ 0.17 & 6.80 $\pm$ 3.11 & 1 \\
3137 &  & 10.52 $\pm$ 0.24 &  &  & 2 \\
4293 & 2.58 $\pm$ 0.05\tablefootmark{a} & 9.30 $\pm$ 0.32 & 1.19 $\pm$ 0.11 & 5.90 $\pm$ 1.82 & 1 \\
4587 &  & 10.91 $\pm$ 0.28 &  &  & 2 \\
8102 &  & 0.79 $\pm$ 0.00\tablefootmark{b} &  &  & 3 \\
 &  & 0.80 $\pm$ 0.00 &  &  & 4 \\
8928 &  & 11.17 $\pm$ 0.15 &  &  & 2 \\
9440 &  & 10.59 $\pm$ 0.32 &  &  & 2 \\
10234 &  & 11.17 $\pm$ 0.21 &  &  & 2 \\
11791 &  & 10.57 $\pm$ 0.17 &  &  & 2 \\
13147 &  & 10.45 $\pm$ 0.17 &  &  & 2 \\
16537 &  & 0.74 $\pm$ 0.01\tablefootmark{b} &  &  & 3 \\
19849 &  & 0.81 $\pm$ 0.00\tablefootmark{b} &  &  & 3 \\
69673 & 1.42 $\pm$ 0.08 &  &  &  & 5 \\
79672 &  & 1.03 $\pm$ 0.01\tablefootmark{b} &  &  & 3 \\
89962 & 3.00 $\pm$ 0.05 &  &  &  & 5 \\
92968 & 2.66 $\pm$ 0.11 &  &  &  & 5 \\
95124 & 3.24 $\pm$ 0.04\tablefootmark{a} & $5.23^{+0.25}_{-0.18}$ & $1.73^{+0.18}_{-0.13}$ & $1.69^{+0.47}_{-0.41}$ & 6 \\
95222 & 2.35 $\pm$ 0.04 &  &  &  & 5 \\
98036 &  & 3.06 $\pm$ 0.02 &  &  & 4 \\
 & 3.53 $\pm$ 0.04 &  &  &  & 5 \\
102014 & 2.57 $\pm$ 0.09\tablefootmark{a} & 10.18 $\pm$ 0.76 & 1.40 $\pm$ 0.22 & 3.40 $\pm$ 1.32 & 1 \\
103836 & 2.66 $\pm$ 0.05\tablefootmark{a} & 7.92 $\pm$ 0.30 & 1.04 $\pm$ 0.10 & 9.90 $\pm$ 2.59 & 1 \\
105854 & 2.42 $\pm$ 0.04\tablefootmark{c} & 10.34 $\pm$ 0.55\tablefootmark{c} & 1.00 $\pm$ 0.16\tablefootmark{c} & 7.29 $\pm$ 3.06\tablefootmark{c} & 7 \\
 & 2.42 $\pm$ 0.07\tablefootmark{a} & 10.85 $\pm$ 0.65 & 1.13 $\pm$ 0.13 & 8.50 $\pm$ 3.33 & 1 \\
110813 & 3.29 $\pm$ 0.04\tablefootmark{a} & $4.53^{+0.13}_{-0.13}$ & $1.46^{+0.09}_{-0.09}$ & $2.46^{+0.67}_{-0.50}$ & 6 \\
 & 3.26 $\pm$ 0.01 & 4.61 $\pm$ 0.09 & 1.42 $\pm$ 0.07 & 2.90 $\pm$ 0.47 & 7 \\
114775 & 2.67 $\pm$ 0.10\tablefootmark{a} & 9.00 $\pm$ 0.65 & 1.38 $\pm$ 0.23 & 4.50 $\pm$ 1.55 & 1 \\
116630 & 3.21 $\pm$ 0.05 & 4.34 $\pm$ 0.21 & 1.12 $\pm$ 0.12 & 7.40 $\pm$ 2.70 & 8 \\
\hline
\end{tabular}
\tablebib{(1): \citet{Aguirre2020}\tablefootmark{d}; (2): \citet{Gallenne2018}; (3): \citet{Tsantaki2013}; (4): \citet{Rains2020}; (5): \citet{Morel2014}; (6): \citet{North2017}; (7): \citet{Campante2019}; (8): \citet{Jiang2020}.
}
\tablefoot{
\tablefoottext{a}{Surface gravity was not listed in the literature, but were computed from the given masses and radius.}\\
\tablefoottext{b}{Radius were not given in \citet{Tsantaki2013}, but were computed from the temperature, angular size, and distance.}\\
\tablefoottext{c}{We used the values corresponding to the Red Clump solution for this star, as it is the most favourable solution based on the arguments presented by the authors.}\\
\tablefoottext{d}{Uncertainties where computed by adding both values listed in \citet{Aguirre2020}.}\\}
\end{table*}

\begin{sidewaystable*}
\small
\caption{SPECIES results for the EXPRESS star sample. Only 40 lines are shown, with a subset of columns. Full version is available online.}\label{tab:expressfull}
\centering
\begin{tabular}{llcccccccccccc}
\hline\hline
Starname & Instrument & Temperature & logg & [Fe/H] & $\xi_t$ & vsini & Mass & Radius & log$\,$L & Age & EEP & P$_{\text{RGB}}$\tablefootmark{a} & P$_{\text{HB}}$\tablefootmark{b} \\
&& (K) & (cgs) & (dex)& (\kms) & (\kms) & ($M_{\odot}$) & ($R_{\odot}$) & ($L_{\odot}$) & (Gyr) & & & \\
\hline
HIP100062 & FEROS & 4933 $\pm$ 50 & 2.63 $\pm$ 0.09 & -0.09 $\pm$ 0.06 & 1.44 $\pm$ 0.06 & 3.80 $\pm$ 0.55 & 1.86$^{+0.10}_{-0.13}$ & 10.75$^{+0.26}_{-0.29}$ & 1.770$^{+0.013}_{-0.014}$ & 1.36$^{+0.22}_{-0.13}$ & 658$^{+6}_{-3}$ & 0.01 & 0.99 \\
HIP101477 & FEROS & 4956 $\pm$ 50 & 2.70 $\pm$ 0.11 & -0.05 $\pm$ 0.06 & 1.50 $\pm$ 0.08 & 3.79 $\pm$ 0.57 & 2.08$^{+0.20}_{-0.20}$ & 10.48$^{+0.38}_{-0.31}$ & 1.770$^{+0.014}_{-0.014}$ & 1.18$^{+0.17}_{-0.28}$ & 668$^{+3}_{-10}$ & 0.01 & 0.99 \\
HIP10164 & FEROS & 4871 $\pm$ 52 & 3.39 $\pm$ 0.10 & 0.12 $\pm$ 0.06 & 0.98 $\pm$ 0.13 & 3.17 $\pm$ 0.49 & 1.19$^{+0.07}_{-0.07}$ & 3.29$^{+0.03}_{-0.03}$ & 0.740$^{+0.009}_{-0.007}$ & 6.05$^{+1.32}_{-1.05}$ & 482$^{+1}_{-1}$ & 1.00 & 0.00 \\
HIP101911 & FEROS & 4864 $\pm$ 50 & 2.85 $\pm$ 0.10 & -0.05 $\pm$ 0.06 & 1.25 $\pm$ 0.09 & 3.36 $\pm$ 0.61 & 1.29$^{+0.10}_{-0.10}$ & 6.22$^{+0.11}_{-0.10}$ & 1.270$^{+0.010}_{-0.011}$ & 4.09$^{+1.16}_{-0.85}$ & 503$^{+1}_{-1}$ & 1.00 & 0.00 \\
HIP102014 & CHIRON & 4677 $\pm$ 50 & 2.71 $\pm$ 0.11 & -0.07 $\pm$ 0.05 & 1.32 $\pm$ 0.09 & 3.68 $\pm$ 0.53 & 1.48$^{+0.12}_{-0.14}$ & 9.89$^{+0.32}_{-0.29}$ & 1.630$^{+0.015}_{-0.016}$ & 2.61$^{+0.96}_{-0.58}$ & 519$^{+1}_{-1}$ & 1.00 & 0.00 \\
HIP10234 & FEROS & 5002 $\pm$ 50 & 2.64 $\pm$ 0.08 & -0.16 $\pm$ 0.05 & 1.46 $\pm$ 0.05 & 3.17 $\pm$ 0.65 & 2.00$^{+0.19}_{-0.17}$ & 10.57$^{+0.32}_{-0.29}$ & 1.790$^{+0.013}_{-0.012}$ & 1.19$^{+0.15}_{-0.23}$ & 665$^{+5}_{-7}$ & 0.00 & 1.00 \\
HIP102773 & CHIRON & 4698 $\pm$ 50 & 2.58 $\pm$ 0.10 & -0.09 $\pm$ 0.06 & 1.52 $\pm$ 0.07 & 4.38 $\pm$ 0.41 & 1.78$^{+0.25}_{-0.19}$ & 13.30$^{+0.52}_{-0.39}$ & 1.900$^{+0.017}_{-0.013}$ & 1.53$^{+0.57}_{-0.51}$ & 681$^{+7}_{-149}$ & 0.39 & 0.61 \\
HIP10326 & CHIRON & 4922 $\pm$ 50 & 2.47 $\pm$ 0.11 & -0.15 $\pm$ 0.06 & 1.42 $\pm$ 0.09 & 3.74 $\pm$ 0.59 & 1.63$^{+0.18}_{-0.15}$ & 11.36$^{+0.25}_{-0.20}$ & 1.820$^{+0.013}_{-0.013}$ & 1.84$^{+0.57}_{-0.42}$ & 668$^{+2}_{-3}$ & 0.00 & 1.00 \\
HIP103836 & CHIRON & 4636 $\pm$ 50 & 2.75 $\pm$ 0.09 & -0.15 $\pm$ 0.05 & 1.18 $\pm$ 0.07 & 3.72 $\pm$ 0.51 & 1.03$^{+0.09}_{-0.07}$ & 7.84$^{+0.21}_{-0.20}$ & 1.420$^{+0.013}_{-0.014}$ & 8.79$^{+2.32}_{-2.04}$ & 512$^{+1}_{-1}$ & 1.00 & 0.00 \\
HIP104148 & FEROS & 4808 $\pm$ 55 & 2.56 $\pm$ 0.12 & -0.05 $\pm$ 0.06 & 1.47 $\pm$ 0.10 & 5.02 $\pm$ 0.44 & 1.65$^{+0.13}_{-0.12}$ & 9.39$^{+0.31}_{-0.32}$ & 1.620$^{+0.020}_{-0.021}$ & 1.78$^{+0.47}_{-0.36}$ & 517$^{+1}_{-1}$ & 0.94 & 0.06 \\
HIP104838 & FEROS & 4911 $\pm$ 50 & 3.20 $\pm$ 0.09 & 0.00 $\pm$ 0.05 & 1.12 $\pm$ 0.09 & 2.71 $\pm$ 0.60 & 1.16$^{+0.08}_{-0.08}$ & 3.84$^{+0.04}_{-0.04}$ & 0.880$^{+0.009}_{-0.007}$ & 6.00$^{+1.51}_{-1.22}$ & 487$^{+1}_{-1}$ & 1.00 & 0.00 \\
HIP10548 & FEROS & 4993 $\pm$ 50 & 3.51 $\pm$ 0.11 & 0.01 $\pm$ 0.05 & 1.23 $\pm$ 0.10 & 2.87 $\pm$ 0.47 & 1.22$^{+0.07}_{-0.07}$ & 3.14$^{+0.06}_{-0.06}$ & 0.730$^{+0.018}_{-0.018}$ & 5.12$^{+1.09}_{-0.79}$ & 480$^{+1}_{-1}$ & 1.00 & 0.00 \\
HIP105854 & FEROS & 4687 $\pm$ 88 & 2.46 $\pm$ 0.16 & 0.16 $\pm$ 0.06 & 1.49 $\pm$ 0.16 & 4.76 $\pm$ 0.55 & 0.97$^{+0.09}_{-0.04}$ & 10.75$^{+0.22}_{-0.16}$ & 1.660$^{+0.011}_{-0.008}$ & 11.50$^{+1.40}_{-2.58}$ & 661$^{+3}_{-4}$ & 0.12 & 0.88 \\
HIP105856 & FEROS & 4801 $\pm$ 62 & 3.05 $\pm$ 0.13 & 0.06 $\pm$ 0.05 & 1.07 $\pm$ 0.14 & 2.41 $\pm$ 0.46 & 1.29$^{+0.10}_{-0.09}$ & 4.99$^{+0.09}_{-0.09}$ & 1.090$^{+0.010}_{-0.010}$ & 4.40$^{+1.26}_{-0.99}$ & 495$^{+1}_{-1}$ & 1.00 & 0.00 \\
HIP106055 & FEROS & 4778 $\pm$ 65 & 2.45 $\pm$ 0.14 & -0.02 $\pm$ 0.06 & 1.39 $\pm$ 0.12 & 3.94 $\pm$ 0.73 & 1.17$^{+0.13}_{-0.13}$ & 11.05$^{+0.12}_{-0.12}$ & 1.730$^{+0.012}_{-0.012}$ & 5.69$^{+2.56}_{-1.52}$ & 664$^{+2}_{-3}$ & 0.00 & 1.00 \\
HIP106922 & FEROS & 4803 $\pm$ 53 & 2.90 $\pm$ 0.11 & 0.01 $\pm$ 0.06 & 1.20 $\pm$ 0.10 & 3.34 $\pm$ 0.58 & 1.27$^{+0.10}_{-0.09}$ & 5.86$^{+0.08}_{-0.08}$ & 1.220$^{+0.010}_{-0.009}$ & 4.52$^{+1.16}_{-0.94}$ & 501$^{+1}_{-1}$ & 1.00 & 0.00 \\
HIP107122 & FEROS & 4965 $\pm$ 50 & 3.28 $\pm$ 0.09 & 0.06 $\pm$ 0.06 & 1.09 $\pm$ 0.09 & 2.88 $\pm$ 0.43 & 1.30$^{+0.08}_{-0.08}$ & 3.80$^{+0.03}_{-0.04}$ & 0.880$^{+0.009}_{-0.008}$ & 4.22$^{+0.93}_{-0.71}$ & 486$^{+1}_{-1}$ & 1.00 & 0.00 \\
HIP107773 & FEROS & 4942 $\pm$ 50 & 2.65 $\pm$ 0.11 & -0.02 $\pm$ 0.06 & 1.47 $\pm$ 0.09 & 4.36 $\pm$ 0.56 & 2.26$^{+0.16}_{-0.24}$ & 11.20$^{+0.29}_{-0.30}$ & 1.830$^{+0.013}_{-0.013}$ & 0.97$^{+0.26}_{-0.20}$ & 676$^{+2}_{-3}$ & 0.01 & 0.99 \\
HIP108543 & FEROS & 5080 $\pm$ 50 & 2.50 $\pm$ 0.12 & 0.06 $\pm$ 0.06 & 1.64 $\pm$ 0.11 & 4.44 $\pm$ 0.59 & 3.03$^{+0.08}_{-0.09}$ & 16.26$^{+0.44}_{-0.41}$ & 2.170$^{+0.018}_{-0.018}$ & 0.40$^{+0.04}_{-0.04}$ & 681$^{+3}_{-4}$ & 0.01 & 0.99 \\
HIP109228 & FEROS & 4993 $\pm$ 50 & 3.42 $\pm$ 0.07 & -0.02 $\pm$ 0.05 & 1.02 $\pm$ 0.08 & 2.28 $\pm$ 0.56 & 1.19$^{+0.07}_{-0.07}$ & 3.21$^{+0.03}_{-0.03}$ & 0.750$^{+0.009}_{-0.007}$ & 5.47$^{+1.07}_{-0.92}$ & 481$^{+1}_{-1}$ & 1.00 & 0.00 \\
HIP110391 & FEROS & 4738 $\pm$ 50 & 2.54 $\pm$ 0.10 & -0.23 $\pm$ 0.06 & 1.35 $\pm$ 0.07 & 3.25 $\pm$ 0.68 & 1.08$^{+0.11}_{-0.10}$ & 9.02$^{+0.28}_{-0.26}$ & 1.540$^{+0.013}_{-0.014}$ & 6.94$^{+2.90}_{-1.92}$ & 516$^{+2}_{-1}$ & 1.00 & 0.00 \\
HIP110529 & FEROS & 5113 $\pm$ 50 & 3.06 $\pm$ 0.11 & 0.09 $\pm$ 0.06 & 1.43 $\pm$ 0.09 & 3.63 $\pm$ 0.60 & 2.21$^{+0.11}_{-0.08}$ & 8.84$^{+0.14}_{-0.24}$ & 1.640$^{+0.012}_{-0.013}$ & 0.89$^{+0.13}_{-0.11}$ & 654$^{+2}_{-12}$ & 0.16 & 0.84 \\
HIP111515 & CHIRON & 4981 $\pm$ 50 & 2.85 $\pm$ 0.11 & 0.03 $\pm$ 0.06 & 1.37 $\pm$ 0.10 & 4.23 $\pm$ 0.49 & 2.08$^{+0.03}_{-0.02}$ & 8.32$^{+0.07}_{-0.06}$ & 1.570$^{+0.007}_{-0.004}$ & 1.03$^{+0.06}_{-0.03}$ & 643$^{+1}_{-1}$ & 0.12 & 0.88 \\
HIP111909 & FEROS & 4920 $\pm$ 54 & 3.11 $\pm$ 0.11 & 0.11 $\pm$ 0.06 & 1.19 $\pm$ 0.11 & 3.20 $\pm$ 0.66 & 1.36$^{+0.09}_{-0.09}$ & 4.29$^{+0.04}_{-0.04}$ & 0.980$^{+0.010}_{-0.009}$ & 3.77$^{+0.97}_{-0.72}$ & 490$^{+1}_{-1}$ & 1.00 & 0.00 \\
HIP113779 & FEROS & 5032 $\pm$ 50 & 3.55 $\pm$ 0.09 & 0.17 $\pm$ 0.05 & 1.00 $\pm$ 0.11 & 2.32 $\pm$ 0.54 & 1.50$^{+0.05}_{-0.06}$ & 3.57$^{+0.03}_{-0.03}$ & 0.850$^{+0.009}_{-0.009}$ & 2.73$^{+0.33}_{-0.24}$ & 483$^{+1}_{-1}$ & 1.00 & 0.00 \\
HIP114408 & CHIRON & 4875 $\pm$ 50 & 3.28 $\pm$ 0.07 & -0.27 $\pm$ 0.05 & 0.92 $\pm$ 0.07 & 3.60 $\pm$ 0.47 & 1.07$^{+0.07}_{-0.07}$ & 4.69$^{+0.04}_{-0.04}$ & 1.060$^{+0.007}_{-0.009}$ & 6.98$^{+1.65}_{-1.38}$ & 495$^{+1}_{-1}$ & 1.00 & 0.00 \\
HIP114775 & CHIRON & 4668 $\pm$ 57 & 2.70 $\pm$ 0.12 & 0.10 $\pm$ 0.05 & 1.22 $\pm$ 0.12 & 4.48 $\pm$ 0.54 & 1.43$^{+0.13}_{-0.12}$ & 8.62$^{+0.27}_{-0.25}$ & 1.500$^{+0.013}_{-0.014}$ & 3.26$^{+1.16}_{-0.78}$ & 514$^{+1}_{-1}$ & 1.00 & 0.00 \\
HIP114933 & FEROS & 4823 $\pm$ 59 & 2.99 $\pm$ 0.12 & 0.06 $\pm$ 0.06 & 1.17 $\pm$ 0.12 & 3.33 $\pm$ 0.51 & 1.39$^{+0.09}_{-0.09}$ & 5.27$^{+0.05}_{-0.05}$ & 1.140$^{+0.009}_{-0.008}$ & 3.40$^{+0.78}_{-0.62}$ & 497$^{+1}_{-1}$ & 1.00 & 0.00 \\
HIP115620 & CHIRON & 4816 $\pm$ 50 & 2.77 $\pm$ 0.09 & -0.00 $\pm$ 0.06 & 1.42 $\pm$ 0.07 & 4.13 $\pm$ 0.50 & 1.90$^{+0.06}_{-0.09}$ & 10.47$^{+0.30}_{-0.25}$ & 1.730$^{+0.014}_{-0.014}$ & 1.36$^{+0.13}_{-0.10}$ & 653$^{+8}_{-7}$ & 0.00 & 1.00 \\
HIP115769 & CHIRON & 4961 $\pm$ 50 & 2.62 $\pm$ 0.12 & -0.26 $\pm$ 0.06 & 1.58 $\pm$ 0.08 & 4.01 $\pm$ 0.56 & 1.24$^{+0.60}_{-0.15}$ & 10.80$^{+0.15}_{-0.26}$ & 1.770$^{+0.009}_{-0.013}$ & 3.94$^{+1.98}_{-2.68}$ & 664$^{+3}_{-6}$ & 0.04 & 0.96 \\
HIP11600 & FEROS & 4969 $\pm$ 64 & 3.69 $\pm$ 0.14 & 0.18 $\pm$ 0.06 & 1.23 $\pm$ 0.10 & 2.89 $\pm$ 0.75 & 1.07$^{+0.04}_{-0.04}$ & 2.43$^{+0.02}_{-0.02}$ & 0.480$^{+0.006}_{-0.009}$ & 9.17$^{+1.27}_{-0.97}$ & 474$^{+1}_{-1}$ & 1.00 & 0.00 \\
HIP116630 & FEROS & 4878 $\pm$ 60 & 3.17 $\pm$ 0.12 & 0.08 $\pm$ 0.06 & 1.18 $\pm$ 0.12 & 2.99 $\pm$ 0.53 & 1.28$^{+0.09}_{-0.09}$ & 4.28$^{+0.05}_{-0.05}$ & 0.970$^{+0.009}_{-0.008}$ & 4.54$^{+1.20}_{-0.92}$ & 490$^{+1}_{-1}$ & 1.00 & 0.00 \\
HIP117314 & CHIRON & 4827 $\pm$ 50 & 2.39 $\pm$ 0.11 & -0.05 $\pm$ 0.06 & 1.49 $\pm$ 0.10 & 4.21 $\pm$ 0.59 & 1.23$^{+0.16}_{-0.12}$ & 11.04$^{+0.12}_{-0.15}$ & 1.750$^{+0.011}_{-0.011}$ & 4.53$^{+1.60}_{-1.33}$ & 665$^{+2}_{-3}$ & 0.00 & 1.00 \\
HIP117411 & FEROS & 4828 $\pm$ 59 & 2.97 $\pm$ 0.12 & 0.06 $\pm$ 0.06 & 1.24 $\pm$ 0.12 & 3.12 $\pm$ 0.61 & 1.18$^{+0.09}_{-0.09}$ & 5.10$^{+0.07}_{-0.07}$ & 1.090$^{+0.010}_{-0.010}$ & 6.08$^{+1.69}_{-1.26}$ & 497$^{+1}_{-1}$ & 1.00 & 0.00 \\
HIP11791 & FEROS & 4958 $\pm$ 50 & 2.75 $\pm$ 0.10 & -0.01 $\pm$ 0.06 & 1.42 $\pm$ 0.08 & 3.24 $\pm$ 0.50 & 2.18$^{+0.15}_{-0.16}$ & 9.90$^{+0.27}_{-0.24}$ & 1.730$^{+0.014}_{-0.014}$ & 1.03$^{+0.20}_{-0.23}$ & 665$^{+3}_{-5}$ & 0.00 & 1.00 \\
HIP11867 & CHIRON & 4775 $\pm$ 51 & 2.55 $\pm$ 0.11 & -0.02 $\pm$ 0.05 & 1.36 $\pm$ 0.09 & 4.67 $\pm$ 0.57 & 1.64$^{+0.16}_{-0.13}$ & 11.84$^{+0.32}_{-0.29}$ & 1.820$^{+0.014}_{-0.013}$ & 2.01$^{+0.51}_{-0.42}$ & 671$^{+2}_{-2}$ & 0.02 & 0.98 \\
HIP1230 & FEROS & 4866 $\pm$ 51 & 2.87 $\pm$ 0.11 & -0.13 $\pm$ 0.05 & 1.28 $\pm$ 0.09 & 3.67 $\pm$ 0.59 & 1.31$^{+0.10}_{-0.10}$ & 6.73$^{+0.10}_{-0.10}$ & 1.340$^{+0.012}_{-0.012}$ & 3.79$^{+1.03}_{-0.75}$ & 506$^{+1}_{-1}$ & 1.00 & 0.00 \\
HIP13147 & FEROS & 4888 $\pm$ 50 & 2.52 $\pm$ 0.07 & -0.36 $\pm$ 0.05 & 1.50 $\pm$ 0.04 & 3.35 $\pm$ 0.62 & 0.89$^{+0.03}_{-0.03}$ & 10.08$^{+0.13}_{-0.11}$ & 1.730$^{+0.012}_{-0.009}$ & 10.09$^{+1.18}_{-1.14}$ & 667$^{+1}_{-1}$ & 0.00 & 1.00 \\
HIP16142 & CHIRON & 4752 $\pm$ 50 & 2.68 $\pm$ 0.10 & 0.05 $\pm$ 0.06 & 1.46 $\pm$ 0.09 & 4.24 $\pm$ 0.54 & 1.74$^{+0.13}_{-0.34}$ & 10.96$^{+0.23}_{-0.39}$ & 1.740$^{+0.011}_{-0.014}$ & 1.72$^{+1.58}_{-0.30}$ & 661$^{+2}_{-10}$ & 0.04 & 0.96 \\
HIP16780 & CHIRON & 5097 $\pm$ 50 & 2.80 $\pm$ 0.07 & -0.26 $\pm$ 0.05 & 1.47 $\pm$ 0.04 & 4.03 $\pm$ 0.51 & 2.21$^{+0.13}_{-0.14}$ & 9.93$^{+0.21}_{-0.19}$ & 1.780$^{+0.015}_{-0.014}$ & 0.84$^{+0.19}_{-0.17}$ & 669$^{+1}_{-3}$ & 0.00 & 1.00 \\
\hline
\end{tabular}
\tablefoot{
\tablefoottext{a}{Probability of the star being in the RGB.}
\tablefoottext{b}{Probability of the star being in the HB.}}
\end{sidewaystable*}

\begin{sidewaystable*}
\caption{Results for the Sun spectra. Spectra labelled Sun01 to Sun07 were taken from \citet{BlancoCuaresma2014}, while the rest were obtained through the ESO data archive.
ATLAS refers to data from \citet{Hinkle2000}, NARVAL to \citet{Auriere2003}, and UVES to \citet{Dekker2000}.}
\label{tab:sunall}
\centering
\begin{tabular}{llcccccccccc}
\hline\hline
Starname & Instrument & [Fe/H] & Temperature & logg & $\xi_t$ & v$\,\sin\, i$ & v$_{\text{mac}}$ & Mass & Radius & $\log\,$L & Age \\
& & (dex) & (K) & (cgs) & (\kms) & (\kms) & (\kms) & ($M_{\odot}$) & ($R_{\odot}$) & ($L_{\odot}$) & (Gyr)\\
\hline
Sun0 & ATLAS & -0.04 $\pm$ 0.05 & 5790 $\pm$ 50 & 4.40 $\pm$ 0.04 & 0.84 $\pm$ 0.04 & 2.82 $\pm$ 0.23 & 3.17 $\pm$ 0.22 & $0.97^{+0.03}_{-0.03}$ & $1.02^{+0.05}_{-0.04}$ & $0.03^{+0.04}_{-0.04}$ & $5.94^{+1.57}_{-1.71}$ \\
Sun1 & HARPS & -0.03 $\pm$ 0.05 & 5776 $\pm$ 50 & 4.38 $\pm$ 0.06 & 0.82 $\pm$ 0.06 & 3.51 $\pm$ 0.17 & 3.21 $\pm$ 0.16 & $0.97^{+0.03}_{-0.03}$ & $1.05^{+0.07}_{-0.06}$ & $0.04^{+0.06}_{-0.06}$ & $6.64^{+1.76}_{-2.18}$ \\
Sun2 & HARPS & -0.02 $\pm$ 0.05 & 5766 $\pm$ 50 & 4.34 $\pm$ 0.04 & 0.86 $\pm$ 0.05 & 3.61 $\pm$ 0.16 & 3.25 $\pm$ 0.16 & $0.97^{+0.03}_{-0.03}$ & $1.10^{+0.06}_{-0.05}$ & $0.08^{+0.05}_{-0.05}$ & $8.01^{+1.40}_{-1.46}$ \\
Sun3 & HARPS & -0.03 $\pm$ 0.05 & 5765 $\pm$ 50 & 4.35 $\pm$ 0.05 & 0.88 $\pm$ 0.05 & 3.83 $\pm$ 0.23 & 3.18 $\pm$ 0.23 & $0.97^{+0.03}_{-0.03}$ & $1.08^{+0.06}_{-0.06}$ & $0.06^{+0.06}_{-0.05}$ & $7.73^{+1.55}_{-1.56}$ \\
Sun4 & HARPS & -0.03 $\pm$ 0.05 & 5767 $\pm$ 50 & 4.36 $\pm$ 0.05 & 0.86 $\pm$ 0.05 & 3.60 $\pm$ 0.16 & 3.21 $\pm$ 0.16 & $0.97^{+0.03}_{-0.03}$ & $1.07^{+0.06}_{-0.06}$ & $0.05^{+0.05}_{-0.05}$ & $7.34^{+1.56}_{-1.66}$ \\
Sun5 & NARVAL & -0.05 $\pm$ 0.05 & 5759 $\pm$ 50 & 4.45 $\pm$ 0.09 & 0.84 $\pm$ 0.10 & 3.77 $\pm$ 0.28 & 2.97 $\pm$ 0.26 & $0.96^{+0.03}_{-0.03}$ & $0.97^{+0.09}_{-0.06}$ & $-0.03^{+0.08}_{-0.06}$ & $4.76^{+2.96}_{-2.95}$ \\
Sun6 & NARVAL & -0.02 $\pm$ 0.05 & 5780 $\pm$ 50 & 4.38 $\pm$ 0.05 & 0.84 $\pm$ 0.05 & 3.42 $\pm$ 0.17 & 3.23 $\pm$ 0.16 & $0.97^{+0.03}_{-0.03}$ & $1.05^{+0.07}_{-0.06}$ & $0.04^{+0.06}_{-0.05}$ & $6.65^{+1.68}_{-1.74}$ \\
Sun7 & UVES & -0.03 $\pm$ 0.05 & 5794 $\pm$ 50 & 4.40 $\pm$ 0.05 & 0.87 $\pm$ 0.05 & 3.14 $\pm$ 0.24 & 3.18 $\pm$ 0.23 & $0.98^{+0.03}_{-0.03}$ & $1.02^{+0.06}_{-0.05}$ & $0.03^{+0.05}_{-0.05}$ & $5.81^{+1.67}_{-1.99}$ \\
ceres01 & HARPS & -0.03 $\pm$ 0.05 & 5757 $\pm$ 50 & 4.33 $\pm$ 0.08 & 0.78 $\pm$ 0.10 & 3.56 $\pm$ 0.26 & 3.18 $\pm$ 0.25 & $0.97^{+0.03}_{-0.03}$ & $1.08^{+0.10}_{-0.09}$ & $0.06^{+0.08}_{-0.08}$ & $7.96^{+1.63}_{-2.44}$ \\
ceres02 & HARPS & -0.02 $\pm$ 0.05 & 5783 $\pm$ 50 & 4.41 $\pm$ 0.05 & 0.77 $\pm$ 0.06 & 3.54 $\pm$ 0.16 & 3.19 $\pm$ 0.16 & $0.98^{+0.03}_{-0.03}$ & $1.02^{+0.06}_{-0.06}$ & $0.02^{+0.05}_{-0.05}$ & $5.74^{+1.76}_{-2.08}$ \\
ceres03 & HARPS & -0.04 $\pm$ 0.05 & 5756 $\pm$ 50 & 4.34 $\pm$ 0.06 & 0.81 $\pm$ 0.07 & 3.43 $\pm$ 0.24 & 3.16 $\pm$ 0.24 & $0.96^{+0.03}_{-0.03}$ & $1.08^{+0.08}_{-0.07}$ & $0.06^{+0.07}_{-0.06}$ & $7.96^{+1.58}_{-1.99}$ \\
moon & HARPS & -0.02 $\pm$ 0.05 & 5761 $\pm$ 50 & 4.39 $\pm$ 0.06 & 0.81 $\pm$ 0.06 & 3.92 $\pm$ 0.23 & 3.08 $\pm$ 0.23 & $0.97^{+0.03}_{-0.03}$ & $1.02^{+0.07}_{-0.06}$ & $0.02^{+0.06}_{-0.06}$ & $6.37^{+1.89}_{-2.29}$ \\
ganymede & HARPS & -0.03 $\pm$ 0.05 & 5777 $\pm$ 50 & 4.36 $\pm$ 0.05 & 0.86 $\pm$ 0.05 & 3.59 $\pm$ 0.16 & 3.24 $\pm$ 0.16 & $0.97^{+0.03}_{-0.03}$ & $1.07^{+0.06}_{-0.05}$ & $0.06^{+0.05}_{-0.05}$ & $7.14^{+1.49}_{-1.69}$ \\
sun01 & HARPS & -0.04 $\pm$ 0.05 & 5752 $\pm$ 50 & 4.32 $\pm$ 0.05 & 0.68 $\pm$ 0.08 & 3.70 $\pm$ 0.23 & 3.17 $\pm$ 0.23 & $0.96^{+0.03}_{-0.03}$ & $1.10^{+0.07}_{-0.07}$ & $0.08^{+0.06}_{-0.06}$ & $8.54^{+1.43}_{-1.65}$ \\
sun02 & HARPS & -0.00 $\pm$ 0.05 & 5893 $\pm$ 64 & 4.47 $\pm$ 0.12 & 0.65 $\pm$ 0.19 & 3.61 $\pm$ 0.35 & 3.44 $\pm$ 0.35 & $1.02^{+0.03}_{-0.03}$ & $1.01^{+0.11}_{-0.06}$ & $0.05^{+0.09}_{-0.07}$ & $3.23^{+2.64}_{-2.10}$ \\
sun03 & HARPS & -0.04 $\pm$ 0.05 & 5791 $\pm$ 50 & 4.36 $\pm$ 0.07 & 0.65 $\pm$ 0.10 & 3.74 $\pm$ 0.24 & 3.25 $\pm$ 0.24 & $0.98^{+0.03}_{-0.03}$ & $1.06^{+0.09}_{-0.07}$ & $0.05^{+0.07}_{-0.07}$ & $6.71^{+1.84}_{-2.40}$ \\
sun04 & HARPS & -0.04 $\pm$ 0.05 & 5743 $\pm$ 50 & 4.25 $\pm$ 0.07 & 0.75 $\pm$ 0.09 & 3.68 $\pm$ 0.24 & 3.28 $\pm$ 0.24 & $0.96^{+0.03}_{-0.03}$ & $1.19^{+0.10}_{-0.09}$ & $0.14^{+0.08}_{-0.07}$ & $9.64^{+1.27}_{-1.32}$ \\
sun05 & HARPS & -0.04 $\pm$ 0.05 & 5772 $\pm$ 50 & 4.36 $\pm$ 0.08 & 0.65 $\pm$ 0.12 & 3.68 $\pm$ 0.25 & 3.19 $\pm$ 0.25 & $0.97^{+0.03}_{-0.03}$ & $1.07^{+0.09}_{-0.08}$ & $0.05^{+0.08}_{-0.07}$ & $7.42^{+1.75}_{-2.64}$ \\
\hline
\end{tabular}
\end{sidewaystable*}

   \begin{figure*}
   \centering
            \includegraphics[width=\textwidth]{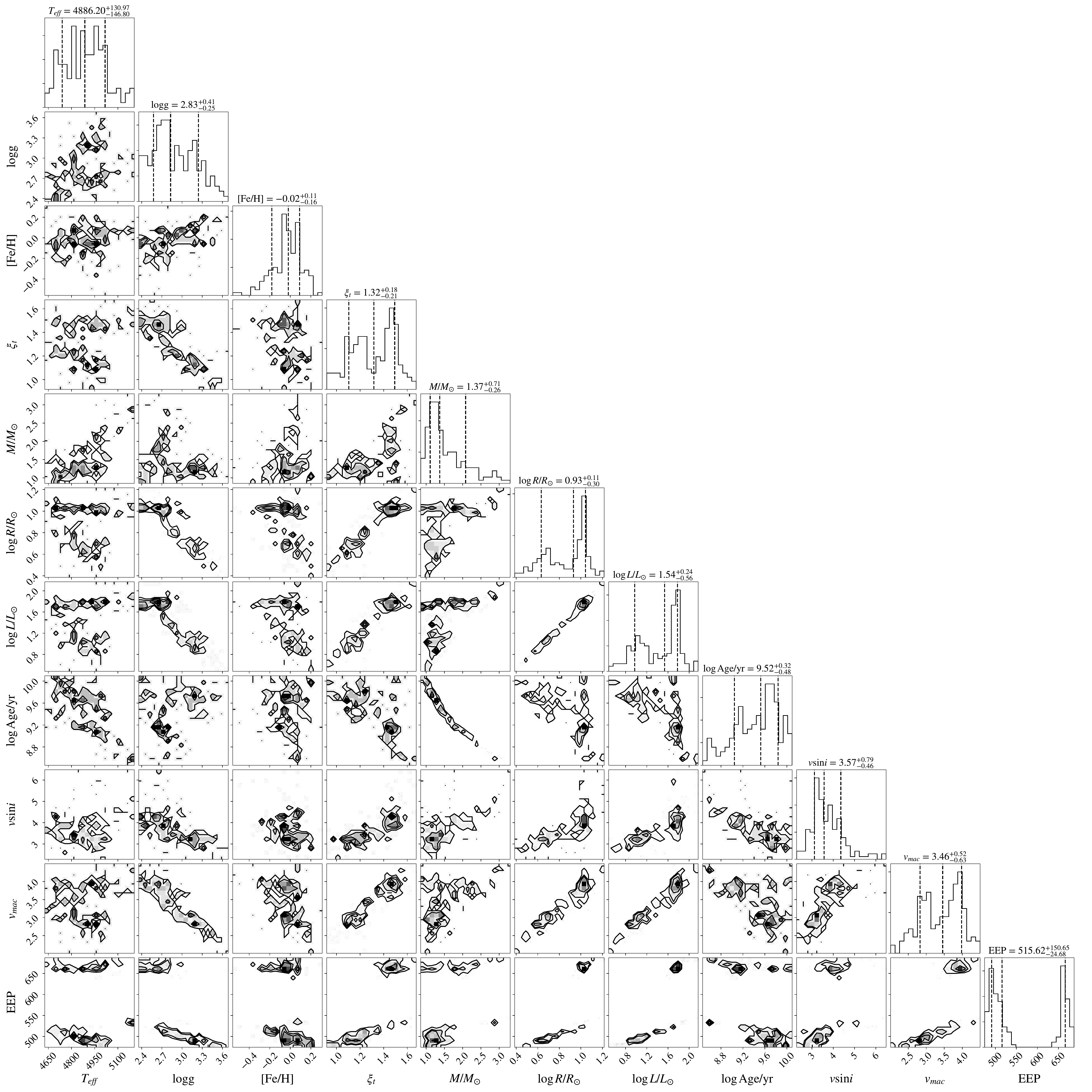}
      \caption{Correlations between all the parameters computed by SPECIES. Only a few of these are plotted in Fig.~\ref{fig:correlations}.
      This figure was generated using the final parameter values for all the stars in the EXPRESS sample.}
         \label{fig:correlations_all}
   \end{figure*}

\begin{table*}
\caption{Pearson correlation coefficient between all the parameters estimated by SPECIES, using the final values computed for the EXPRESS sample.}
\label{tab:correlations}
\centering
\begin{tabular}{l | rrrrrrrrrr}
 & $T_{\text{eff}}$ & logg & [Fe/H] & $\xi_t$ & $\log\, M/M_{\odot}$ & $\log\, R/R_{\odot}$ & $\log\, L/L_{\odot}$ & $\log\,$Age/yr & $v\sin i$ & $v_{mac}$ \\
 & (K) & (cgs) & (dex) & (\kms) & & & & & (\kms) & (\kms) \\
\hline
logg & 0.314 &  &  &  &  &  &  &  &  &  \\
{[Fe/H]} & 0.109 & 0.409 &  &  &  &  &  &  &  &  \\
$\xi_t$ & 0.172 & -0.783 & -0.167 &  &  &  &  &  &  &  \\
$\log\, M/M_{\odot}$ & 0.654 & -0.234 & 0.186 & 0.632 &  &  &  &  &  &  \\
$\log\, R/R_{\odot}$ & 0.016 & -0.891 & -0.306 & 0.903 & 0.589 &  &  &  &  &  \\
$\log\, L/L_{\odot}$ & 0.116 & -0.852 & -0.299 & 0.911 & 0.657 & 0.994 &  &  &  &  \\
$\log\,$Age/yr & -0.654 & 0.268 & -0.101 & -0.652 & -0.993 & -0.621 & -0.689 &  &  &  \\
$v\sin i$ & 0.231 & -0.536 & -0.086 & 0.728 & 0.607 & 0.71 & 0.726 & -0.626 &  &  \\
$v_{mac}$ & 0.181 & -0.876 & -0.354 & 0.903 & 0.579 & 0.934 & 0.944 & -0.614 & 0.681 &  \\
EEP & 0.067 & -0.711 & -0.173 & 0.779 & 0.514 & 0.819 & 0.822 & -0.523 & 0.505 & 0.763 \\
\hline
\end{tabular}
\end{table*}

\end{appendix}


\end{document}